\documentclass[12pt,preprint]{aastex}

\slugcomment{To appear in the Astronomical Journal}

\shorttitle{NGC 2264}
\shortauthors{Sung et al.}

\begin{document}

\title{A {\it Spitzer} View of the Young Open Cluster NGC 2264}

\author{Hwankyung Sung\footnote{Visiting Associate, Spitzer Science Center,
 California Institute of Technology, 314-6, Pasadena, CA 91125}}
\affil{Department of Astronomy and Space Science, Sejong University,
    Kunja-dong 98, Kwangjin-gu, Seoul 143-747, Korea}
\email{sungh@sejong.ac.kr}

\author{John R. Stauffer}
\affil{Spitzer Science Center, California Institute of Technology,
314-6, Pasadena, CA 91125}
\email{stauffer@ipac.caltech.edu}

\and

\author{Michael S. Bessell}
\affil{Research School of Astronomy and Astrophysics, Australian
National University, MSO, Cotter Road, Weston, ACT 2611, Australia}
\email{bessell@mso.anu.edu.au}

\begin{abstract}
We have performed mid-IR photometry of the young open cluster NGC 2264 using
the images obtained with the {\it Spitzer} Space Telescope IRAC and MIPS
instruments and present a normalized classification scheme of young stellar objects
in various color-color diagrams to make full use of the information from
multicolor photometry. These results are compared with the classification
scheme based on the slope of the spectral energy distribution (SED).

From the spatial distributions of Class I and II stars, we have identified
two subclusterings of Class I objects in the CONE region of Sung et al.
The disked stars in the other star forming region S MON are mostly
Class II objects. These three regions show a distinct difference in
the fractional distribution of SED slopes as well as the mean value of
SED slopes. The fraction of stars with primordial disks is nearly flat
between $\log m =$ 0.2 -- -0.5, and that of transition disks is very high for
solar mass stars.
In addition, we have derived a somewhat higher value of the primordial
disk fraction for NGC 2264 members located below the main pre-main sequence
locus (so-called BMS stars).  This result
supports the idea that BMS stars are young stars with nearly edge-on disks.
We have also found that the fraction of primordial disks is very low near
the most massive star S Mon and increases with distance from S Mon.

\end{abstract}

\keywords{stars: formation -- stars: pre-main sequence -- planetary systems:
protoplanetary disks -- open clusters and associations: individual (NGC 2264)}

\section{INTRODUCTION}

Curiosity about the formation of stars and planets is fundamental to mankind.
The hypothesis of planet formation in the protosolar
nebula can be traced back to the 18th century. Despite good theoretical support
for the existence of disks around young stars \citep{lbp74}, 
they were largely ignored for many years because the existing observational
data could be explained without reference to disks \citep{klt07}.
This situation changed slowly over time, particularly as infrared
detectors became sensitive to the warm dust component of circumstellar
disks \citep{ssecs, lw84}.
The 2 Micron All Sky Survey (2MASS - \citet{2mass}) provided an unprecedented
amount of data in the near-IR passbands that were used
in many important studies, such as the distribution of embedded star forming
regions (SFRs), statistical estimates of young stellar objects (YSOs),
and the general distribution of stars in the Galaxy. 
However in the earliest stages of star formation, 
the spectral energy distributions are dominated by radiation from circumstellar
dust at mid- and far-infrared wavelengths, with little at near-IR wavelengths of 2MASS.

The launch of the {\it Spitzer} Space Telescope revolutionized 
the study of star formation. IRAC (Infrared Array Camera - \citep{irac}) and 
MIPS (Multiband Imaging Photometer for {\it Spitzer} - \citep{mips}) instrument
teams and several Legacy projects obtained observations of young open clusters
and individual young SFRs, including ``From Molecular Cores to Planet-Forming
Cores (c2d)'' \citep{nje03}, ``Formation and Evolution of Planetary Systems
(FEPS)'' \citep{jmc08}, ``Taurus - The Spitzer Map of the Taurus Molecular
Clouds'' \citep{dlp07}, and ``Gould's Belt - Star Formation in the Solar
Neighborhood'' \citep{lea07a} - providing a greatly improved census and
characterization  of young stars and their disks in a wide range of
star-forming environments. 

As an example of the breadth of these surveys, the c2d project aimed to
survey with IRAC and MIPS, five nearby star-forming clouds within a certain
reddening boundary, and to produce catalogs and spectral energy distributions
(SEDs) for YSO members over the entire SFR (i.e. not spatially biased to
only the cloud cores). IRAC and MIPS instrument teams surveyed relatively
distant young open clusters and OB associations under GTO programs, e.g. 
IC 348 \citep{cjl06}, NGC 2264 \citep{pst06,ety06}, $\sigma$ Ori \citep{hj07a}, 
Cep OB2 \citep{saa06}, etc.
Many scientific papers have already been published from these projects.
Amongst them, \citet{ph07} studied the YSOs in the Serpens
cloud and found that disks in less evolved stages were more clustered.
\citet{rag08} also found a different spatial distribution of Class I and
II objects in NGC 1333. Class I objects in NGC 1333 were located along
the filament, while Class II objects were more widely distributed. 
Recently \citet{lc07} studied the disk properties of a specific
type of stars - weak-line T Tauri Stars (WTTSs). Many investigators have
previously derived the overall fraction of primordial disks\footnote{
The primordial disk is defined as the disks around a YSO resulted from
the star formation process - Class I and II objects} around YSOs in young
open clusters and groups \citep{dh07,hj08,mhlf05}, and tried to determine
the lifetime of primordial disks to contrain the planet-building time scale.
Although many attempts have been made to determine the effects of massive stars
on disk evolution, little evidence has been found.

The core cluster of the Mon OB1 association, NGC 2264, is one of the most
studied young open clusters. Although the cluster is more distant than
the Orion nebula cluster (ONC),
NGC 2264 is still relatively nearby and serves as a good prototypical
populous, young open cluster (d $\approx$ 760 pc:
\citet{sbl97}). In addition, the large number (about 1000) of
known pre-main sequence (PMS) members and nearly zero foreground reddening
has attracted many observational studies from radio to X-ray wavelengths. 
In this section we will summarize the most relevant studies only
(see \citet{sed08} for an extensive review of NGC 2264).
\citet{daa72} discovered a strong mid-IR source in NGC 2264 (IRS1).
Later, \citet{hch77} observed this object with the NASA G. P. Kuiper Airborne
Observatory in the far-IR and suggested that it might be a stellar or
protostellar object of a moderate mass, $\sim$ 5 -- 10 M$_\odot$.
\citet{mly89} studied the Mon OB1 (NGC 2264) molecular clouds with IRAS
12 $\mu m$, 25 $\mu m$, 60 $\mu m$ and 100 $\mu m$ images and found
30 discrete far-IR sources. They also calculated the spectral index of
these sources and found that 18 IRAS sources were Class I objects. They
concluded that star formation is ongoing in NGC 2264. From the $K$ band
luminosity function \citet{lyg93} estimated the number of cluster members
to be 360 $\pm$ 130. \citet{lmr02} performed an optical and near-IR survey
of the cluster, presenting photometry for over 5600 stars. They used various
disk and mass accretion indicators and derived an inner disk fraction ranging
from 21\% to 56\% depending on which empirical disk indicator was used,
and a typical mass accretion rate on the order of
10$^{-8}$ M$_\odot$ yr$^{-1}$. Recently, \citet{pst06}, using mid-IR images
obtained with the {\it Spitzer} Space Telescope, found a dense embedded
mini-cluster (Spokes cluster) around the strong IR source IRS2 in NGC 2264.
They found uniformly spaced protostars in a semi-linear fashion.
Based on
the fact that the distance between the nearest protostars was of the order of
the Jeans length, they concluded that this geometry was evidence for thermal
fragmentation. \citet{ety06} studied the SED
of objects around IRS2 and confirmed the youth of these 
objects. \citet{sbcki08} published a vast amount of optical photometric
data for NGC 2264. From the surface density distribution of H$\alpha$
emission stars, they identified two dense SFRs (S MON and CONE) and a HALO
region surrounding the two SFRs.

In this paper, we present the overall properties of NGC 2264 in the mid-IR
passbands. The data from this work will be used in a future paper on
the initial mass function and young brown dwarf candidates
in NGC 2264. We will describe the photometry from {\it Spitzer} images
in \S 2. Along with the mid-IR photometric data for about 22,000 objects,
optical and 2MASS counterparts of these mid-IR sources will be identified.
In addition the X-ray and 
H$\alpha$ emission stars in mid-IR color-color (C-C) diagrams will
be presented in this section. The classification scheme used in this study
is described and compared with previous methods in \S 3. 
The mid-IR color-magnitude diagrams (CMDs) and the optical CMD of stars
detected in the mid-IR photometry will also be presented in this section.
In \S 4 we will identify two young sub-clusterings including the
Spokes cluster from the spatial distribution of Class I sources and show
the fractional distribution of the SED slope
of YSOs in each sub-clustering. The fraction of disked stars will be
derived in \S 5 and the effect of massive stars on primordial disks will
be examined in \S 6. And the disk fraction of below the PMS locus (so-called
BMS) stars will be discussed in \S 7. The mid-IR characteristics of
several interesting objects are also discussed in this section.
The summary of this work is in \S 8.

\section{OBSERVATION AND DATA REDUCTION}

\subsection{{\it Spitzer} Observations of NGC 2264}

The {\it Spitzer} mapping observations were performed under program
ID 37 (PI: G. Fazio) in 7 $\times$ 11 mosaics. Each pointing was imaged in
the high dynamic range mode (exposure time: 0.4 sec and 10.4 sec).
The mapping of NGC 2264 was performed on 2004 March 6 and October 8, with
a different spacecraft position angle at the two epochs. 
Two iterations each day were made with an offset
of about 12 arcseconds. The observed region is slightly larger than the
HALO region defined in \citet{sbcki08}.  The AORs utilized for these maps
were numbers 3956480, 3956736, 3956992, and 3957248.

MIPS scanning of NGC 2264 was obtained on 2004 March 16 (PID 58) at the medium
scan rate (exposure time : 3.67 s). Fourteen scans of 0.75$^\circ$ length,
with 160$''$ offsets
have were used. The observed area is larger and fully covered the region
observed with IRAC.   The AOR utilized for the MIPS mapping was number 4317184.
The Post-BCD (basic calibrated and mosaiced) images were downloaded from
the {\it Spitzer} archive. The pixel size of the IRAC Post-BCD data is 
$1.''2 \times 1.''2$, while that of the MIPS 24 $\mu m$ data is 
$2.''45 \times 2.''45$.   The data utilized  pipeline processing software
version S14.0.0 for the IRAC images and S16.1.0 for the MIPS 24$\mu m$ image.

\subsection{Photometry}

We have used the IRAF\footnote{IRAF (Image Reduction and Analysis Facility) is developed
and distributed by the National Optical Observatories.} version of DAOPHOT
to derive PSF-fitting photometry for stars in the field of NGC 2264.
Because {\it Spitzer} IRAC images
are undersampled, point spread function (PSF) fitting yields photometry
with relatively poor signal-to-noise. For uncrowded fields with little
nebulosity, aperture photometry would provide photometry with lower noise
than PSF-fitting photometry for IRAC data.  However, portions of the
NGC 2264 field are crowded or have highly variable and strong nebulosity -
or both. We believe that the PSF-fitting photometry provides a more
uniform and reliable photometry for the entire set of cluster stars,
admittedly at the expense of having more noise than aperture photometry
for the stars where the backgrounds are benign and crowding is not an issue.
For most stars in the cluster, because we have four independent sets of
data, the PSF-fitting photometric accuracy is improved by averaging
the results from the separate AORs.

To minimize the contribution of the spatially varying nebulosity, we used a 
very small FIT radii (1.4 pixels) and then corrected to 3-pixel radii 
magnitudes using the magnitude differences at the two radii calculated from 
bright stars free of nebulosity. The sky background was taken 
from the annulus between 3 and 7 pixels from the source.
As Post-BCD images are calibrated, re-sampled and mosaiced data, the spatial
variation inherent in the original BCD images is altered and removed.
For this reason we assumed that there is no spatial variation in the PSF
for Post-BCD images, i.e. set the `varorder' parameter in daopars to ``0''.

The Post-BCD mosaic images are flux calibrated in units of MJy sr$^{-1}$.
The photometric zero points to use for the PHOT module of DAOPHOT
were calculated using the pixel scale and the flux
of a zeroth magnitude star from the IRAC Handbook. The values used were 
17.30, 16.82, 16.33 and 15.69 mags for channels 1 through 4, respectively. 
The aperture corrections
(to an aperture = 10 pixels, sky annulus = 10 -- 20 pixels) applied were
-0.127, -0.130, -0.145 and -0.228 mags for channels 1 through 4, respectively.
\citet{jlh08} gave slightly different aperture corrections for post-BCD
images, but the differences are less than 0.015 mag which has a negligible 
effect on determining physical properties for our stars.
As MIPS 24 $\mu m$ images are well sampled data, PSF photometry for them is
rather straightforward. We used a 2.4-pixel FIT radii and 20$''$ -- 32$''$
for the sky annuli. The Post-BCD images were calibrated in units of MJy
sr$^{-1}$ and was 141.1 $\mu$Jy pix$^{-2}$ for MIPS 24$\mu m$ images.
The photometric zero point for use with PHOT (11.76 mag) was calculated using the pixel
scale and flux of a zeroth magnitude star in the MIPS handbook (7.14 Jy). 
The final 24 $\mu$m magnitudes [24] were obtained by applying an aperture 
correction of -0.52 mag.

It is worth mentioning two points relating to the photometry. The first concerns
the detection and removal of cosmic rays. 
Because the IRAC images are undersampled it is very difficult to
distinguish cosmic rays from point sources. 
We used the fact that we have multiple, independent mosaics of the cluster
to eliminate cosmic rays not removed within an individual AOR by the
MOPEX mosaicing software.   That is, we created catalogs of all point-like
sources for each AOR.  When we combined the four catalogs, we rejected all
sources only identified in a single AOR.

The second point concerns point source photometry in spatially varying nebulosity. 
In order to help eliminate false point sources that are in fact knots or
edges in the nebula, we created median-subtracted mosaic images.  We visually
examined these median-subtracted images for all the point sources in our 
master catalog that lie within the nebular regions - and deleted all
sources that do not appear point-like in the median-subtracted images.
Our final photometry was still performed on the original mosaic images, however.

\begin{figure}
\epsscale{0.5}
\plotone{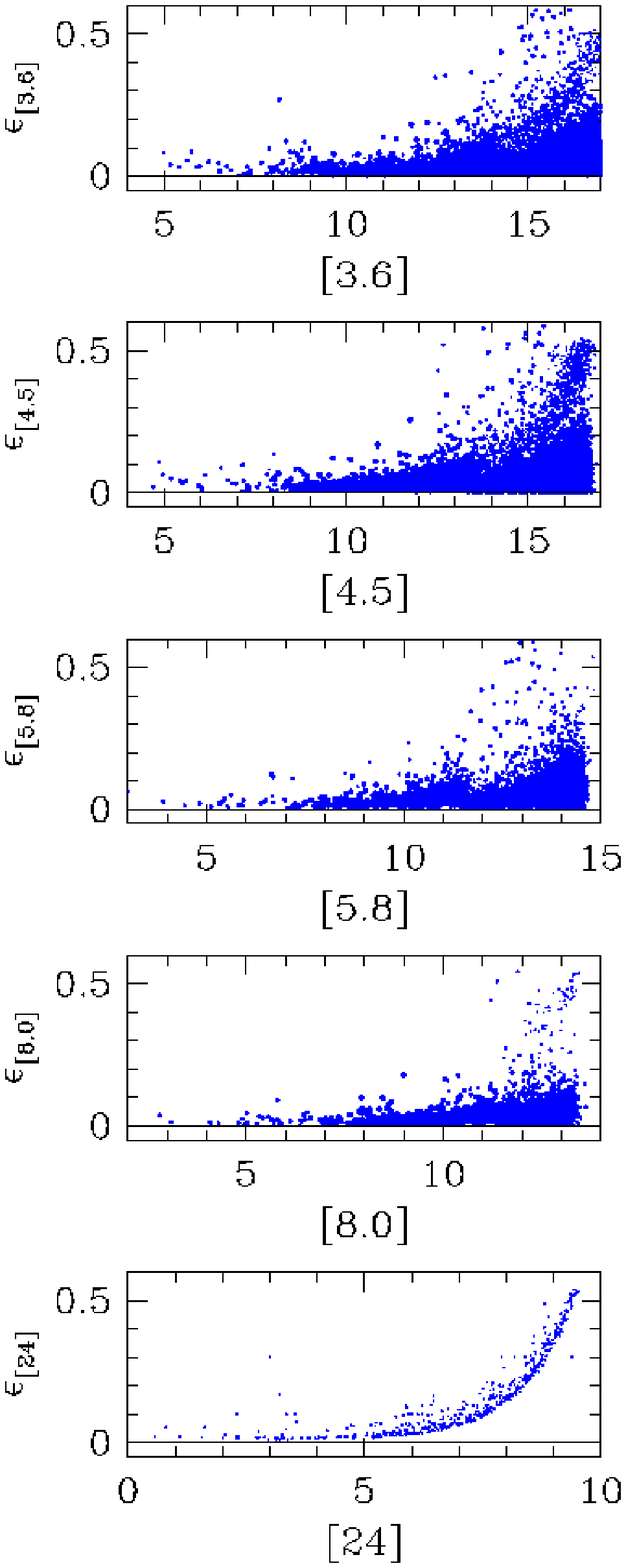}
\caption{The distribution of photometric errors as a function of magnitude.
The size of symbols are proportional to the number of observations.
\label{figerr} }
\end{figure}

We present the photometric data from four IRAC bands and the MIPS 24 $\mu m$
band
for 21,991 objects in Table \ref{tab_sst}. We label the objects in Table
\ref{tab_sst} as SST + the identification number in the first column. 
The total number of objects detected from the photometry are 21224, 18329,
5113, 2302 and 506 in the 3.6 $\mu m$, 4.5 $\mu m$, 
5.8 $\mu m$, 8.0 $\mu m$ and 24 $\mu m$ bands, respectively. 
The weighted mean values and weighted errors of the
magnitude from multiple observations were calculated as in \citet{sl95}
(weight = 1 / $\epsilon^2$). The distribution of photometric errors is
shown in Figure \ref{figerr}. The distribution of photometric errors in [24]
is the typical distribution of errors from PSF photometry of well sampled data.
For undersampled data, like the IRAC or HST/WFPC2 images, the distribution
is far different - the photometric errors are no better than 0.1 mag even
for bright stars (see the error distribution of bright stars detected only
in the short exposure images in Figure \ref{figvar}).
But as 4 epochs of data, with 2 exposure times per epoch are available,
there could be a maximum number of eight detections for some objects.
If the magnitudes from
all images are consistant, the resulting final error is small. If not,
the resulting error will be larger. Such a trend can be seen in Figure
\ref{figerr} -- there are many faint stars with small errors.

Due to the lower spatial resolution of the MIPS 24 $\mu m$ data,
we paid considerable attention to finding the appropriate IRAC counterparts. 
If two or more IRAC sources were matched as counterparts of a [24] source
(matching radii = 2.$''5$), we applied the following rule for the [24]
magnitude. If neither of them showed any evident IRAC excess, then the [24]
flux was assigned to the nearest object (2 cases out of 18). If only one
source showed a certain or probable IRAC excess (9 cases),
then the [24] flux was assigned to that IR excess star. For one case, although both
objects showed an IRAC excess, we assigned the MIPS flux to only one of them
because that source was much brighter than the other.  In four cases more than one
IRAC source of similar brightness 
showed an IR excess. For such cases we drew a wavelength versus
brightness plot, extrapolated to 24 $\mu m$ and then determined the brightness
difference at 24 $\mu m$. Using this difference and measured [24] flux, we 
derived an estimated
[24] mag for each component. In the extrapolation, we assumed a logarithmic
function-like variation for Class II stars and a cubic function-like
variation for Class I objects. For these objects we assigned a large error
-- 0.1 mag if the two objects were of similar brightness or 0.3 mag if the
brightness difference was large (see the lowest panel of Figure \ref{figerr}).
For 2 cases, because both objects are very faint in [3.6] and [4.5] and not
detected in [5.8] and [8.0], the [24] flux was not assigned to either of them.

\placefigure{figerr}
\placetable{tab_sst}

We compared our data with \citet{ety06} and found
no counterparts in our data for 50 of their objects. Most of them were detected in only
one channel and may therefore be spurious detections such as cosmic
ray events.   For the objects in common with \citet{ety06},
the differences relative to our photometry are +0.151 $\pm$ 0.097
(N = 50, 5 excluded), +0.077 $\pm$ 0.127 (N = 50, 4 excluded),
+0.081 $\pm$ 0.137 (N = 42, 3 excluded), +0.195 $\pm$ 0.164 (N = 32,
2 excluded) in [3.6], [4.5], [5.8] and [8.0], respectively.
The systematic offsets relative to the \citet{ety06} data may result from 
their use of aperture photometry (or, at least from the fact that we use
PSF photometry and they use aperture photometry). We also compared our data
with independent aperture photometry kindly determed by I. Song (priv. comm.).
The difference in [3.6] is +0.036 $\pm$ 0.024 (N = 51, 7 excluded).
Recently \citet{cb07} published photometric data for 445 stars in NGC 2264.
They published data in fluxes rather than in magnitudes. We transformed their
data to magnitudes using the photometric zero points in the IRAC Handbook 
(V3.0). The differences relative to our photometry were +0.126 $\pm$
0.052 (N = 408, 18 stars excluded), +0.158 $\pm$ 0.046 (N = 396, 25 stars
excluded), +0.104 $\pm$ 0.064 (N = 312, 40 stars excluded) and +0.154 $\pm$
0.056 (N = 184, 18 stars excluded) in [3.6], [4.5], [5.8] and [8.0],
respectively. The differences are relatively large and systematic, but we cannot
identify the source of the systematic differences. To help 
determine the source of the
differences, we also calculated and compared the $K_s$-[3.6] colors for 
relatively bright, diskless stars in the NGC 2264 field (specifically, stars
with [3.6]-[4.5] $<$ 0.2 and [3.6] $<$ 13). 
The mean $K_s$-[3.6] colors of our data was +0.170 $\pm$ 0.104
(N = 183), while that of \citet{cb07} was +0.294 $\pm$ 0.120 (N = 183).
As the reddening of stars in NGC 2264 is very small \citep{sbl97},
the $K_s$-[3.6] color of Class III objects or photospheres should be close to 0.
We believe that the smaller mean $K_s$-[3.6] color for our photometry 
supports our contention that our systematic difference relative to 
\citet{cb07} and relative to \citet{ety06} is due to a zero-point error
in their photometry, at least for 3.6 $\mu m$.

\subsection{Optical and 2MASS Counterparts, and X-ray sources}

We searched for optical and 2MASS counterparts to the sources detected
in IRAC with a search radius of 1$''$. The master optical source catalogue
we used was created from Tables 3 (``C'' + ID objects), 8 (``W'' + 4 digit ID
objects), and 9 (``S'' + ID objects) of \citet{sbcki08}.
Among 21991 IRAC objects, 7869 objects
have a 2MASS counterpart and 17932 objects have an optical counterpart.
As the optical data had better spatial resolution, 5 IRAC sources had 
3 optical counterparts and 496 sources had two. 
Two active PMS stars (C30987 \& C30995) were located within
the matching circle of SST 10207. The distance between the two stars is about
0.$''$9. A strong X-ray emission object (detection significance = 25.59) was 
also detected at the same position. \citet{ds05} reported a strong H$\alpha$
emission source ($W_{\rm H\alpha}$ = 458.2 $\AA$). Either or both stars 
could be the optical counterpart of SST 10207.
As in most cases Class I objects do not have normal optical colors,
we tentatively identify the faint BMS star C30995 as the optical counterpart
of the Class I star SST 10207. The one Class I star that we believe does
fall in the PMS locus is SST 12918 (= C35527), an X-ray emission star.

\begin{figure}
\epsscale{0.5}
\plotone{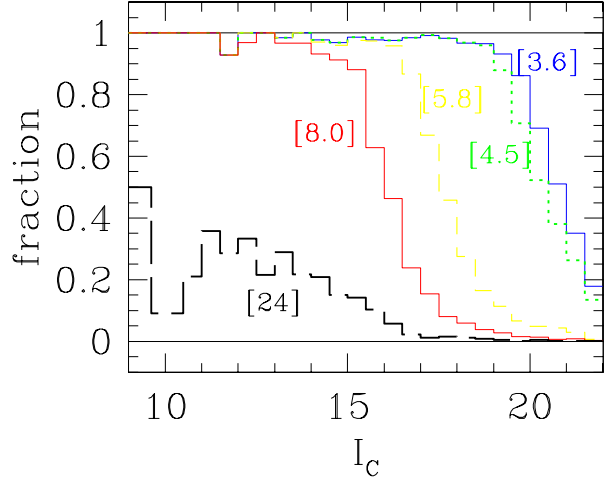}
\caption{The completeness of {\it Spitzer}
photometry for the 16,772 optical stars in the central region of NGC 2264 covered
by all 4 IRAC channels. The 80\% completeness is $I_C$ = 20, 19.5, 17, and 15.5
for [3.6], [4.5], [5.8], and [8.0], respectively. The completeness
of [24] is generally below 40\%. \label{figphotc} }
\end{figure}

To check for completeness relative to the optical photometry, we identified
{\it Spitzer} counterparts of optical sources in the central region of
NGC 2264 covered by all 4 IRAC channels. The area is 1204 arcmin$^2$.
There were 16,772 optical stars, excluding stars where a single {\it Spitzer} source
is matched to two or more optical stars (522 optical stars in total).
The completeness is shown in Figure \ref{figphotc}. The 80\% completeness 
level was $I_C$ = 20, 19.5, 17 and 15.5 for [3.6], [4.5], [5.8] and
[8.0], respectively. Fainter than that, the detection fraction decreases 
rapidly.  Because stellar photospheres are relatively faint
at 24 $\mu m$, the detection fraction in [24] is generally less than 40\%.

\placefigure{figphotc}

In addition, we also searched for any new infrared counterparts of X-ray sources (X-ray
sources from \citet{sbc04,sbcki08} - X-ray sources in \citet{svr04,ef06} were
included). We identified 52 IRAC sources as new counterparts
of X-ray sources with significance $>$ 5 and 30 IRAC sources as counterparts of
weak X-ray sources with significance between 3.3 and 5.0. Among them, SST
13029, rather than C35645 in Table 3 of \citet{sbcki08}, is the actual
counterpart of X-ray source F06-264 (\citet{ef06}, CXOJ0641071+0930366).
Many of the new X-ray counterparts are embedded Class I or II objects. 
Some of them may be unseen active galaxies.  
We list {\it Spitzer} data for the 43 X-ray sources without previous
optical or 2MASS counterparts in Table \ref{tab_Xray}.

\begin{figure}
\epsscale{1.0}
\plotone{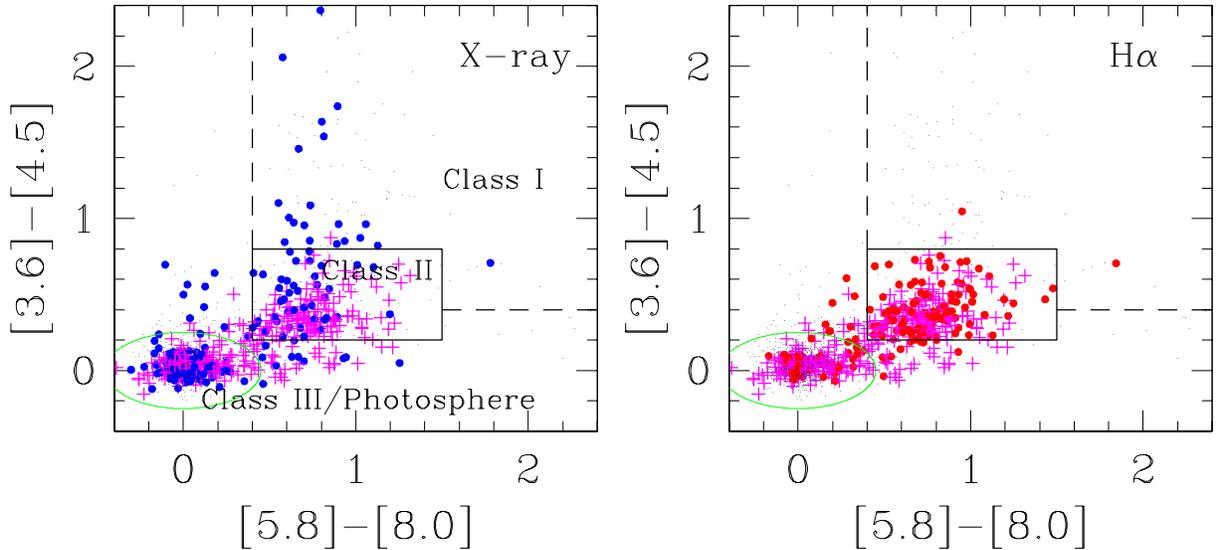}
\caption{The distribution of X-ray emission stars and H$\alpha$ emission
stars in the ([3.6]-[4.5], [5.8]-[8.0]) diagrams.
The large (blue) dots in the left panel represent X-ray emission stars,
while a (magenta) plus sign in both panels denotes a star
showing emission both in X-ray and H$\alpha$. The large (red) dots in the right
panel are H$\alpha$ emission stars. The small dots are stars with no emission
in X-ray (left) or in H$\alpha$ (right). \label{figXnH} }
\end{figure}

The optical and 2MASS data for all {\it Spitzer} sources are listed in Table
\ref{tab_sst}. In addition, other information related to cluster membership
such as H$\alpha$ emission or X-ray emission (both for our new IR counterparts
and for objects previously cross-identified with X-ray sources based on optical
and near-IR data), is also included. Figure 
\ref{figXnH} shows the distribution of all NGC 2264 X-ray emission and
H$\alpha$ emission sources in the ([3.6]-[4.5], [5.8]-[8.0]) diagram. We adopt
the classification scheme of \citet{stm04,lea04,lea07b}.
Large dots represent either X-ray emission stars (left) or H$\alpha$
emission stars (right), while plus marks denote stars with emission in both
X-ray and H$\alpha$. 
Most of the H$\alpha$ emission stars are distributed in the region
occupied by Class II and Class III/Photosphere stars. Only three 
H$\alpha$ stars are in
the locus of Class I objects. On the other hand, X-ray emission objects
are more widely distributed. There are many X-ray sources in the region
occupied by Class I or highly reddened objects.
These very red ([3.6]-[4.5] $>$ 1.5) X-ray emission objects are located 
in the Spokes
cluster or the Cone nebula region, and therefore are almost certainly deeply embedded Class I
objects rather than background active galaxies.

\placetable{tab_Xray}
\placefigure{figXnH}

\subsection{Very Red Objects}

\subsubsection{Objects That Are Red in [5.8]-[8.0]}

There are a total of 38 objects with [5.8]-[8.0] $>$ 1.5 mag. We carefully
examined the long exposure optical images obtained with
the Canada-France-Hawaii Telescope (CFHT) (see \citet{sbcki08}) to help
determine their nature.
Based on our examination of the optical images, we classify
22 of these objects as galaxies -
three are obvious galaxies and 19 appear as
diffuse objects, which we believe are best interpreted as galaxies.
These 22 objects are marked as ``Galaxy'' in the 20th
column of Table \ref{tab_sst}. Three objects are point-sources that
have been identifies as having H$\alpha$ emission. Two of them are 
BMS stars with H$\alpha$ emission - the well-known 
BMS star W90 (=S2144 = SST 9597)
and C31519 (= SST10475). One is a suspected BMS star (C32005 = SST 10710
= Ogura 97) - \citet{ko84} classified the star as an H$\alpha$ emission star,
but the star showed no indication of H$\alpha$ emission from our
H$\alpha$ photometry.
Two other optical counterparts are point sources in the optical without
detected H$\alpha$ emission - both lay below the 
PMS locus in the $I_C$ versus ($R-I$)$_C''$  CMD. 
We could find no optical counterparts for the remaining 11 objects.

\subsubsection{Objects That Are Red in [3.6]-[4.5]}

We also examined the optical images of 25 objects that were very red in
[3.6]-[4.5] ([3.6]-[4.5] $>$ 0.4 mag \& [5.8]-[8.0] $<$ 0.4 mag).
This region in the IRAC C-C plane is known as the locus of reddened Class II
objects \citep{lea04}. We could find no optical counterparts for 20 of the
objects. Most of them are in regions of bright nebulosity. This fact supports
the proposition that objects in this region of the C-C diagram are probably
embedded and therefore highly reddened.
Five objects are detected in the optical and are point sources.
One of them is probably a field star, where the red [3.6] - [4.5] color is
probably not real but due to a large error in the photometry. 
The other four stars have H$\alpha$
emission and are probably YSOs (One is an X-ray emission star, another is a BMS star).

\subsection{Galaxy Contamination}

As mid-IR photons are less affected by interstellar reddening, contamination
of cluster stars by external galaxies is inevitable. The galaxies
with red [5.8]-[8.0] colors mentioned in \S 2.4.1 are bright in [8.0]
due to 7.7 $\mu m$ -- 8.2 $\mu m$ Polycyclic Aromatic Hydrocarbon (PAH)
emission excited by young, high-mass stars \citep{stern05,rag08}.
But as mentioned above, a well-known Herbig Be star W90 (= S2144) also shows
emission in [8.0] (see also \citet{ldk08}). Star S2975 (= SST 12833) and
SST 10184 (= C30920+C30962) show similar characteristics (H$\alpha$ emission
and [8.0] emission).

We constructed SEDs for all of our candidate PMS objects, and identified
all objects where the [8.0] flux is significantly brighter than a curve
drawn through the other IRAC fluxes (extending to the 24 $\mu m$ point where
the object is detected at 24 $\mu m$).  For the objects with significant
flux excess at 8 $\mu m$, we examined our deep optical images at the
coordinate of these objects. Among 8 objects with PAH emission (excluding
the objects already mentioned in \S 2.4.1 and above paragraph), we newly
identified SST 6182 and 19807 as galaxies, but no optical counterparts could
be found for SST 10914, 11099, 13654, 14676, 14997, and 17616.

\citet{rag08} found the locus of star-forming galaxies (SFGs) dominated by
PAH emission in the ([3.6]-[5.8], [4.5]-[8.0]) and ([4.5]-[5.8],
[5.8]-[8.0]) diagrams. There are 22 objects in our NGC 2264 data that
fall within the \citet{rag08} SFG locus in the ([4.5]-[5.8], [5.8]-[8.0])
diagram.  Thirteen of these objects have 8 $\mu$m excesses according to
our SED analysis and were classified as galaxies based on our
examination of the deep CFHT images.  Two more of the objects in
the SFG locus were also clasified by us as galaxies but do not
show 8 $\mu$m excesses.  The remaining seven objects in the SFG
locus are SST  10184, 10914, 11099, 13654, 14676, 14997, and 17616.
 SST 10184 has two optical counterparts (C30920 \& C30962)
within the search radius. If the actual counterpart is C30962, then SST 10184
could be a faint SFG.
As SST 10914 (Class II) and SST 11099 (Class I) are in the two active SFRs
(Spokes and Cone (C) - see \S 4.1), they could be deeply embedded
YSOs. Three objects (SST 14676, 14997, 17616) without optical counterparts
are in the Halo or Cone (H) regions, and therefore they may be SFGs
with PAH emission.  The other two objects have spatial
locations that do not allow us to confidently identify them as
galaxies or YSOs.

The results from the ([3.6]-[5.8], [4.5]-[8.0]) diagram is nearly the same
as above. Only one object identified in the ([4.5]-[5.8], [5.8]-[8.0])
diagram
as a PAH emission SFG, SST 5721 (an optically confirmed galaxy),
was missed in the ([3.6]-[5.8], [4.5]-[8.0]) diagram.

In addition to SFGs with [8.0] PAH emission,
AGNs also show excess emission in the IR. The locus of AGNs overlaps
with that of Class I objects in IR color-color diagrams. \citet{rag08}
filtered out broad-line AGN candidates using color and magnitude cuts
in the [4.5] versus [4.5]-[8.0] diagram (including eliminating all objects
fainter than [4.5] $>$ 14.5). We have attempted a different approach,
by directly examining deep optical images of  all
95 Class I objects selected in \S 3.1. Among them, 68 objects had no optical
counterpart (nothing seen). Most of them were clustered around the Spokes
cluster or the Cone nebula; we believe their spatial location makes it highly
likely these objects are deeply embedded YSOs. Ten objects were point
sources in the deep optical images.
Among the 10 point sources, 5 objects (SST 2558, 10207, 11837, 12918, \&
13288) were surrounded by a thin nebulosity (i.e. the wings of the PSF are
slightly elevated). SST 2558 is an H$\alpha$
emission star (\# 45 of \citet{rpabv04}), and may be an optical counterpart
of {\it ROSAT} X-ray emission object 1WGA J0640.1+0943.   Because of these
properties, and because it is relatively bright ($I_C$ = 17.8 mag), we
believe
it is unlikely that SST 2558 is an external galaxy.  The other four objects
are located either in the Spokes cluster or the Cone nebula, and therefore
we assume they are exposed YSOs with nebula around them. Another 8 objects
were
very faint, and it was difficult to discern whether they were faint stars or
galaxies. We found extended nebulosity at the coordinates of 5 objects
(SST 12208, 12253, 12583(?), 13566 \& 14214), but the nebulae seem to be
nebulae excited by embedded YSOs (that is, the nebulae were structured or
asymmetric, unlike what we would expect for a distant galaxy).
Four of the sources (SST 4787, 6262, 16492 and 21671) were diffuse,
and based on our experience they are most likely galaxies
(They are marked as ``Galaxy'' in
the 20th column of Table \ref{tab_sst}). Because we found several galaxies
close in vicinity to SST 8501, 9861 and 21808, we suspect these sources are
galaxies and members of small groups or clusters.  In total, 7 of the Class I
objects are best categorized as galaxies based on our examination of the deep
optical images; 21 of the Class I objects cannot be verified or rejected; and
67 of the Class I objects are best interpreted as YSOs (because of their
spatial
location or association with asymmetric nebulae).
In support of our analysis, we note that
the visually identified or suspected galaxies are more-or-less randomly
scattered throughout the outer portion of the IRAC field of view (see
Figure \ref{figmap}).

\subsection{Variability}

One of the important characteristics of YSOs is their photometric variability. 
Several attempts have been
made to detect variability in the {\it Spitzer} data. \citet{lmr07,ph07}
searched for variability of YSOs in Serpens or Perseus, but due to the
short time span they could not detect variability above the 10\% or 20\%
level. Recently, \citet{upv08} detected the variability of objects in the Large
Magellanic Cloud from the SAGE (Surveying the Agents of a Galaxy's Evolution)
survey. In addition, \citet{mmc08} obtained the most extensive IRAC monitoring
of a YSO field to date, consisting of twice daily monitoring of the IC1396A
star-forming core. They found that $>$ 1/3 of the YSOs in this $\sim$1 Myr
cluster are variable at IRAC wavelengths, with amplitudes up to 0.2 mag; about
a dozen of the sources appeared to be periodic variables, most with periods
of order 3-10 days, but with one star having a 3.5 hour period (a possible
PMS $\delta$ Scuti star).

\begin{figure}
\epsscale{1.0}
\plotone{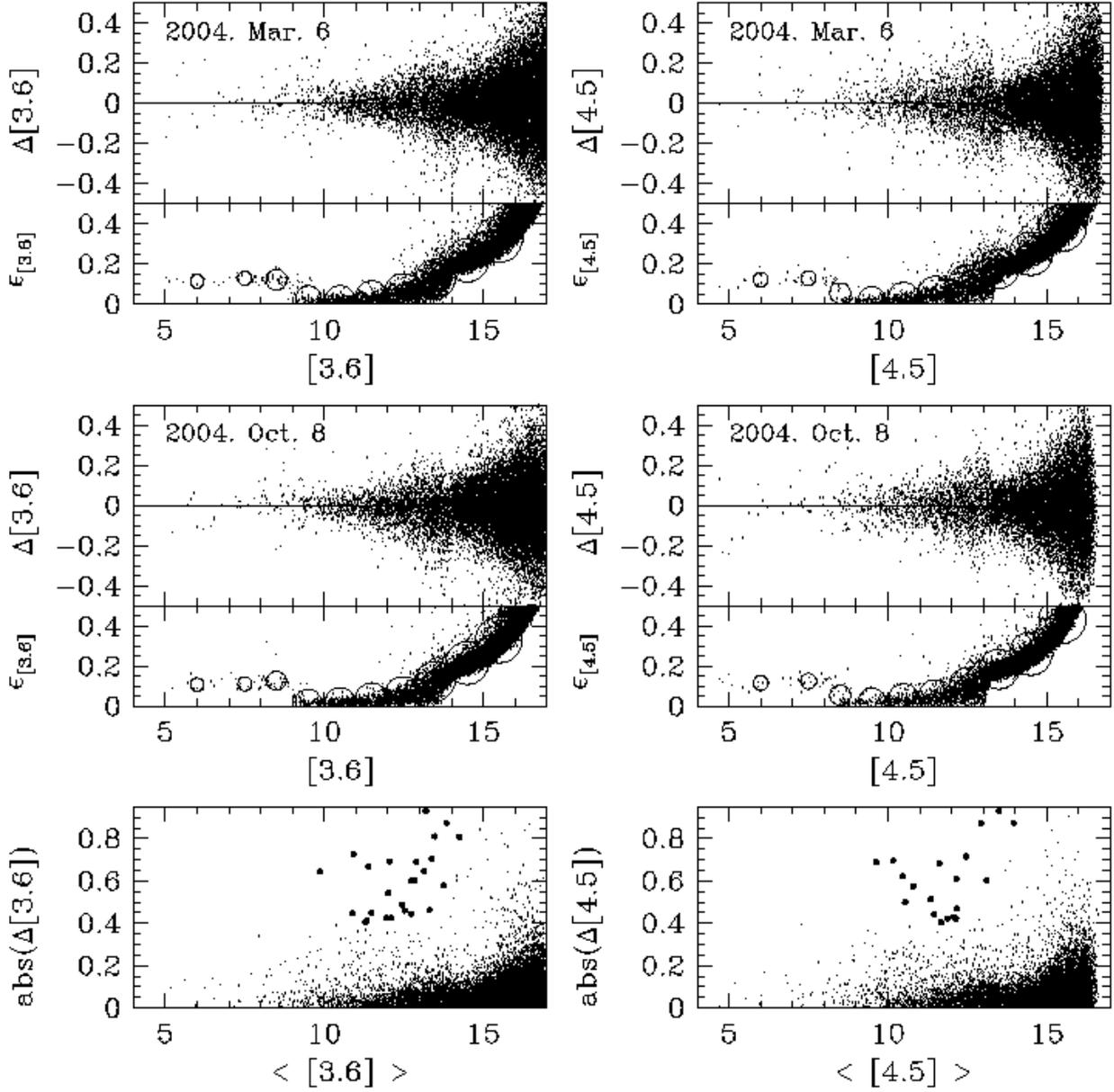}
\caption{Variability test at 3.6 $\mu m$ (left panels) and 4.5 $\mu m$
(right panels). The upper two panels show the magnitude difference and total
error of data obtained on 2004 March 6. Open circles in the second panel
represent the mean errors for a given magnitude bin. The size is proportional
to the number of stars in the bin. The middle two panels show similar plots for
2004 October 8. The lowest panels show the distribution of magnitude
differences between the two epochs. The large red dots denote variable candidates.
\label{figvar} }
\end{figure}

Because we have 4 sets of data observed at two widely separated epochs, it is 
worthwhile to check for variability in our catalogue. \citet{upv08} introduced 
a variability index based upon the error-weighted flux difference between
two epochs in order to search for variables.   We have
searched for variable objects in NGC 2264 in a similar way. First we calculated
the weighted mean and total error (square root of the quadratic sum of 
photometric errors for each set) in [3.6] and [4.5] for data obtained 
on the same date.
We then calculated the differences in magnitude, the ratio of the difference,
and the total error for a given passband. If the difference between two epochs
was greater than 3.5 times the total error and greater than about 0.4 mag -
the photometric error of DAOPHOT undersampled data was about 0.1 mag -
then we classified the object as a variable source. In this process we
selected 28 variable candidates. Figure \ref{figvar} shows the distribution
of brightness difference between two data sets obtained on the same date 
(uppermost and middle panel) as well as between two widely separated epochs 
in the lowest panel. 
The left panel is for [3.6] and the right panel is for [4.5]. The large dots
in the lowest panels represent the variable candidates selected from this
work and listed in Table \ref{tab_var}. The number of sources showing
variability is actually a lower limit because we performed PSF photometry which
gives a larger photometric error for the undersampled data.  A reanalysis
of the data using aperture photometry would undoubtably identify a larger
number of stars that are variable at IRAC wavelengths.

\placefigure{figvar}
\placetable{tab_var}

In total, 28 variable candidates were selected. In general, the variation
pattern and amplitude in [3.6] and [4.5] was very similar. Only two objects
(SST 13432 \& 13612) showed a variation $>$ 0.4 mag in one channel but only
a small ($<$ 0.15 mag) delta magnitude in the other channel - the variations
in these two stars may be spurious.
Among the 28 variable candidates, 4 objects were classified as Class I, 8
were Class II and 3 were intermediate between Class II and III (see \S 3 for
the classification scheme). The others candidates were not classified because
of a lack of either [5.8] or [8.0] data or large errors at the long wavelengths.
But most of the remaining variable candidates 
are very red.  Based on a review of the photometry, we can offer
tentative classifications for five of the remaining stars. Although the photometric
errors are large, SST 11699 and 14514 should probably be
classified as Class I. SST 15726 is probably a Class II and SST 13123 is on
the border between Class I and II. SST 8570 was detected in all IRAC channels
with somewhat larger errors and is a Class III or normal star.
The others were only detected in [3.6] and [4.5] and their [3.6]-[4.5] colors
are red.

\section{CLASSIFICATION OF YOUNG STARS}

\citet{lw84} presented the SEDs of deeply embedded YSOs in $\rho$ Oph. 
Later, \citet{cjl87} introduced the slope of the SED,
$\alpha \equiv {{d \log (\lambda F_\lambda) } \over {d \log \lambda}}$ to
classify YSOs. Now, the slope of the SED is a basic tool with which to diagnose
the evolutionary stage of YSOs. The innovative development of two-dimensional
photodetectors in the infrared, such as IRAC and MIPS 
on the {\it Spitzer} Space Telescope, make it posssible to observe 
many YSOs in a single image;
as a result, the number of YSOs from a single observation can increase from
a few up to several hundred. This large increase in the number of objects 
considered has made photometric methods based on photometric C-C
diagrams a much more powerful way to classify the evolutionary stage of YSOs.
In addition, modeling techniques have also improved \citep{lea04, tpr07},
resulting in better predictions of the locus of YSOs at a given 
evolutionary stage. 

In this section we attempt to classify the stars detected from the {\it Spitzer}
images of NGC 2264 using primarily photometric colors. We also use the SED
method as an ancillary tool.

\subsection{{\it Spitzer} Color-Color Diagrams}

\begin{figure}
\epsscale{1.0}
\plotone{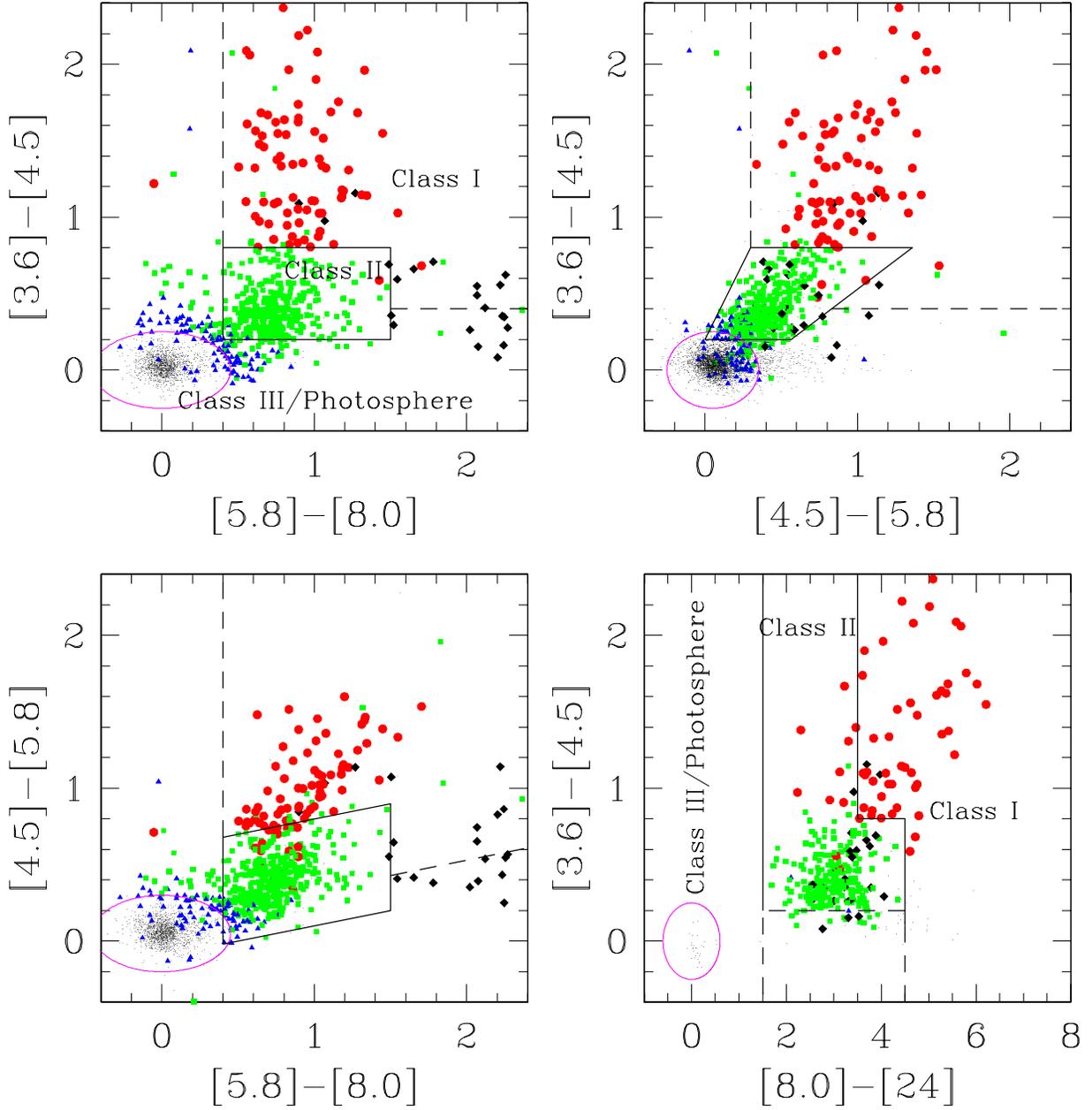}
\caption{The distribution of YSOs in various color-color diagrams. The loci
of Class I and II objects are adopted from \citet{stm04} and \citet{lea07b}.
A large red dot, green square, and blue triangle denotes, respectively, Class I,
II, and II/III (intermediate between Class II and III) objects based on the
weighting scheme described in the text. Black diamonds represent optically
confirmed background galaxies. Small dots are either Class III objects,
stellar photospheres or objects having bad quality data. We assume the locus of
Class III/Photosphere objects is an ellipse. \label{figccd} }
\end{figure}

\citet{stm04} introduced the classification criteria in the ([3.6]-[4.5],
[5.8]-[8.0]) C-C diagram. Later, slight modifications \citep{ph07,rag08}
or the inclusion of MIPS [24] data \citep{lea07b} were made.
In the classification of YSOs in NGC 2264, we basically used the locus of
YSOs introduced by \citet{stm04} and \citet{lea07b} and extended them to
other C-C diagrams. The locus of Class III and stellar-photosphere stars
was assumed to be an ellipse around (0, 0) in the C-C diagram.
In addition, we adopted a weighting scheme in the classification
because in many cases the class of a YSO from one C-C
diagram differs with that from other C-C diagrams.
Two kinds of weights were used - the first ($q_i$ = 1.5, 0.7, 1.0 and 1.5*1.5
for C-C diagrams in Figure \ref{figccd} from upper left, upper
right, lower left and lower right, respectively.) was related to
the classification resolution (eg. the class of a YSO from the ([3.6]-[4.5],
[5.8]-[8.0]) diagram was more reliable than that from the ([3.6]-[4.5],
[4.5]-[5.8]) diagram.). The second ($w_i$) was related to the data quality
(i.e. photometric errors). We assigned $w_i$ = 0.3, 0.7 and 1.5 for bad, fair,
and good data, respectively (`bad' if the individual error of all relevant
colors were greater than 0.1 mag, `good' if the individual error of all relevant
colors were smaller than 0.1 mag and the total error was smaller than 0.2 mag,
and `fair' if between them). As the main purpose of this study was to select
members of the young open cluster NGC 2264, we assume that objects outside
the locus of Class I or Class III/Photosphere are Class II. The mean value
of a YSO class ($Cl_i$) (call this quantity $Q_{CC}$) is then

$$ Q_{CC} \equiv {{\sum_i Cl_i \cdot q_i \cdot w_i} \over {\sum_i q_i \cdot w_i} } $$

\placefigure{figccd}

If $Q_{CC}$ is equal to or less than 1.5, the object is assigned to Class I.
If $Q_{CC}$ is smaller than 2.3, the object is assigned to Class II.
And if $Q_{CC}$ is between 2.3 and 2.7, the object is Class II/III (an object
intermediate between Class II and III). Practically we only classified
objects detected in all four IRAC colors with good quality data (A quadratic
summed total error in all 4 IRAC channels of less than 0.25 mag).

\subsection{Spectral Energy Distribution}

The slope of the SED from IRAC [3.6] to MIPS [24] was calculated
for the objects detected in all 4 IRAC channels. In the calculation of the SED
slope, we used a weighted linear fit to the derived fluxes. The weight applied
was the inverse of the quadratic sum of the photometric errors and the 
uncertainty in the
photometric zero point (assumed to be 0.05 mag). The adopted flux of 0th magnitude
stars and the effective wavelengths from the optical to near-IR are those by
\citet{bcp98}. For {\it Spitzer} passbands, these values were taken from the 
IRAC and MIPS Data Handbooks\footnote{$F_{\nu,0}$ of 0th magnitude star is
280.9 $\pm$ 4.1 Jy, 179.7 $\pm$ 2.6 Jy, 115.0 $\pm$ 1.7 Jy, 64.1 $\pm$ 0.9 Jy,
and 7.14 $\pm$ 0.0815 Jy for IRAC 3.6 $\mu m$, 4.5 $\mu m$, 5.8 $\mu m$, 8.0 $\mu m$,
and MIPS 24 $\mu m$, respectively.} (version 3.0 and 3.3.0, respectively).

To calculate the reddening corrected SEDs, a reddening law should also
be adopted. In general, that for the $UBV$ passbands is well known and agrees
for different objects (see \citet{bcp98} or \citet{elf99}), but those for 
$R_C$ to the mid-IR
differ markedly (see \citet{ri05} or \citet{nlc08}). For consistency,
we adopted the reddening law from \citet{bcp98} for the optical to the 
near-IR, but for
{\it Spitzer} passbands we derived color excess ratios from the slope of
reddened stars in Figure \ref{fig2ms}. The color excess ratios (E($K_s$-[IRAC])
/ E($J-H$)) derived from Figure \ref{fig2ms} are 0.39, 0.47, 0.56 and 0.555
for [3.6], [4.5], [5.8] and [8.0], respectively. It is not easy to compare
the ratios because $A_{[IRAC]} / A_{K_s}$ are also affected by the adopted
color excess ratios. If we adopt $A_{J,H,K_s} / A_{K_s}$ of \citet{ri05} 
our IR reddening law is similar to dust models of \citet{wd01} with $R_V$ = 3.1.
If we adopt the color excess ratios of \citet{bcp98}, the $A_\lambda / A_{K_s}$
of our IR reddening law is similar to that derived for the case of $A_{K_s} \geq 2$ 
of \citet{nlc08}. Currently we use the former case as we have adopted
the color excess ratios of \citet{bcp98}. The color excess ratio for [24]
shows a large variation (see Table 3 of \citet{nlc08}). We tentatively assume
that $A_{[24]} /A_{K_s}$ is the same as $A_{[8.0]} /A_{K_s}$.

In the classification of the evolutionary stage of YSOs using the slope of the
SED, we adopted the classification scheme of \citet{gwayl94} and
\citet{cjl06}. \citet{cjl06} defined objects with primordial disks as those 
with $\alpha > -1.8$ rather than using the classical \citet{gwayl94} definition
of Class II objects ($\alpha > -1.6$). \citet{dh07} calculated the fraction of stars with
primordial disks using the same Lada et al. criterion and as we would also 
like to calculate the disk fraction and compare it with previous work, we 
will use the same criterion ($\alpha \geq -1.8$) for Class II.

Class I   : $\alpha \geq $ +0.3

flat      : +0.3 $> \alpha \geq$ -0.3

Class II  : -0.3 $> \alpha \geq$ -1.8

Class III : -1.8 $> \alpha \geq$ -2.55

\begin{figure}
\epsscale{1.0}
\plotone{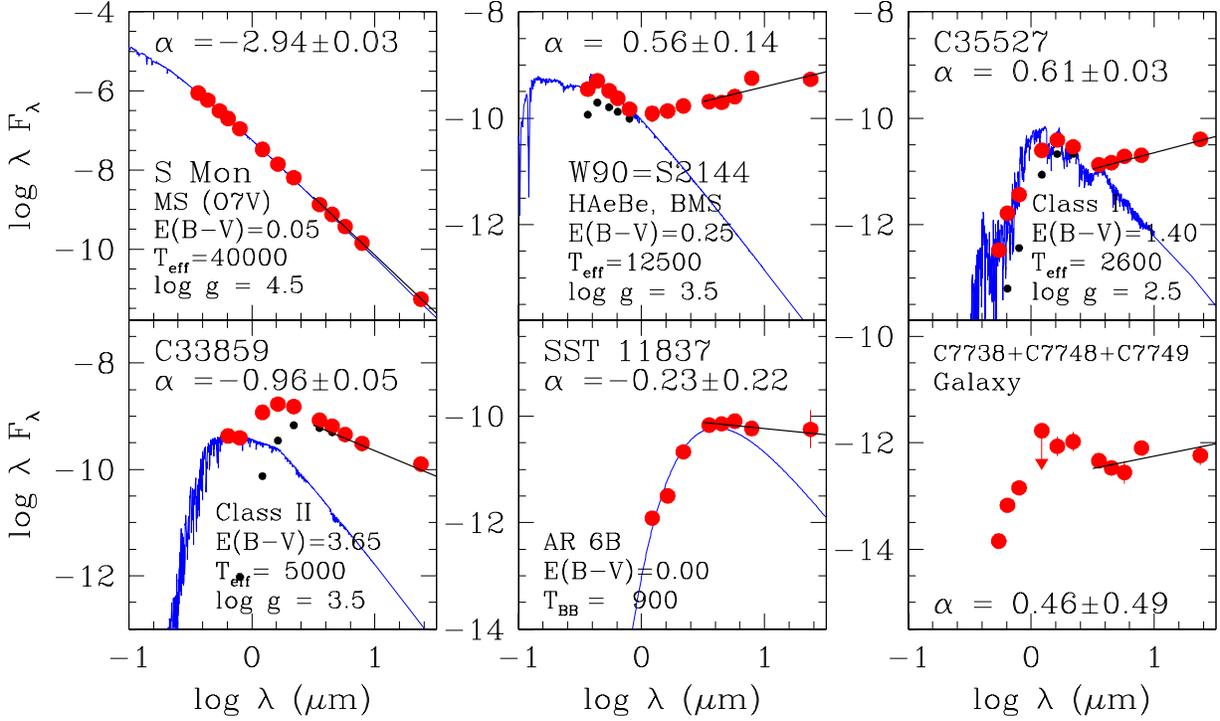}
\caption{Typical Spectral Energy Distributions. The SEDs presented here
are the SED of S Mon (O7V, typical stellar photosphere): the
flux of a Castelli-Kurucz model atmosphere (see \citep{bcp98})
(upper left); of a below the main-sequence Herbig Be star
W90 (= S2144): Castelli-Kurucz flux (upper middle); of a Class
I object in the PMS locus: PMS model atmosphere from NextGen (upper
right); of a FU Ori type candidate AR 6A and its companion AR 6B (lower left
and lower middle); and of a star-forming galaxy (lower right). Optical photometric
data are from \citet{sbcki08} and near-IR data are from 2MASS \citep{2mass} 
except for AR 6B. The near-IR data for AR 6B are from \citet{ar03}.
\label{figsed} }
\end{figure}

\placefigure{figsed}

We present several typical SEDs in Figure \ref{figsed}. The first sample is
the SED for S Mon (O7V), the most massive star in NGC 2264. Ideally the slope
of the SED of early type stars should be -3.0, but due to the uncertainty in
the calibration of 0th magnitude in the mid-IR, the derived slope is 
-2.94 ($\pm$ 0.03). The synthetic spectra superposed is Castelli's Kurucz model
atmosphere (see \citet{bcp98}). The second SED is that of the well-known
BMS Herbig Be star W90. More discussion on W90 is
presented in \S 7.2.1. The third example is the SED of a Class I object in
the PMS locus - C35527 (see \S 7.2.2 for more details). The model atmosphere
superposed is the NextGen model atmosphere \citep{hab99} for PMS stars
(\url{ftp://ftp.hs.uni-hamburg.de/pub/outgoing/phoenix/Pre-MS}). The spectral
resolution of the theoretical fluxes were reduced for purposes of presentation. 
The 2MASS colors of C35527 do not show any signature of excess emission when
we fit the photometry up to $K_s$.  Close examination of this fit suggests
that a cooler model photosphere (than the T$_{eff}$ = 2600K model we derived)
might be more appropriate, but the NextGen models for PMS stars
only provide spectra for temperatures between 2600 -- 6800 K,
As a result the reddening may be overestimated.
The next two SEDs are that of the FU Ori candidate
AR 6A and its companion candidate AR 6B \citep{ar03} (see \S 7.2.3).
The last SED is the SED for a star-forming galaxy (see \S 2.4.1). The galaxy
was measured as three point sources, but the $I_C$ band image shows that they
are three bright regions in a galaxy. The 8 $\mu m$ PAH feature
is prominent in the star-forming galaxies. A more detailed
analysis of YSO SEDs in NGC 2264 will be presented in Bayo et al. (2009, 
in preparation).

\begin{figure}
\epsscale{1.0}
\plotone{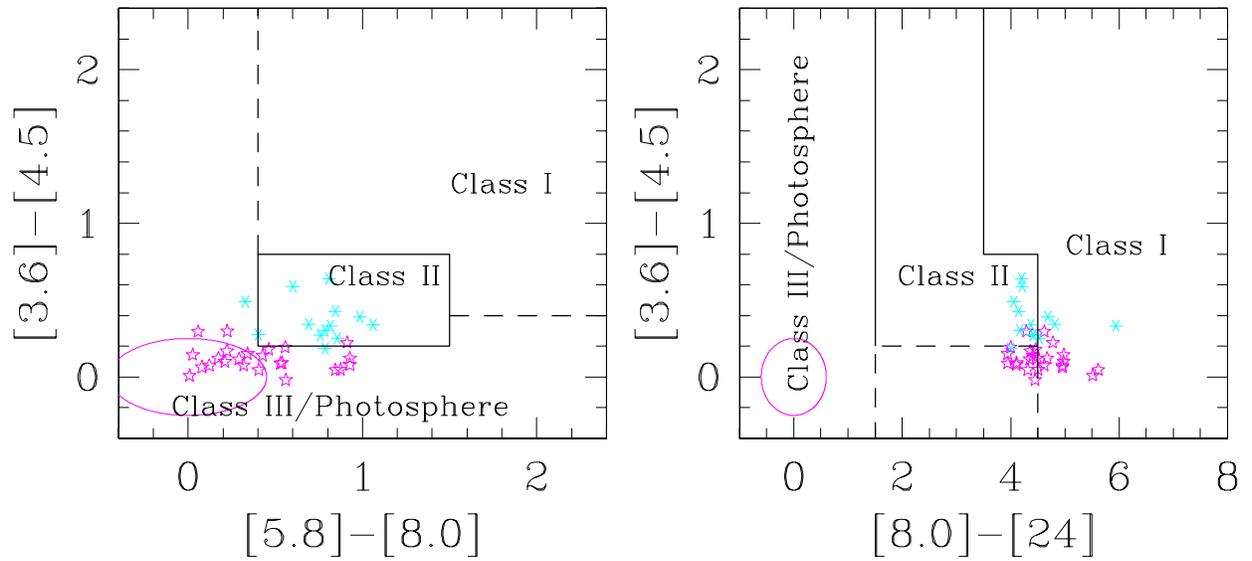}
\caption{Distribution of objects with transition disks (star marks) or with
pre-transition disks (asterisks). In the left panel, objects with transition
disks are more concentrated near the locus of Class III/Photosphere objects,
but on the right they are located around the border between Class I and II.
\label{figtdisk} }
\end{figure}

We also classified objects with pre-transition disks \citep{ce08} or transition
disks. See \citet{nje09} for physical and observational definitions of these
terms. 
\citet{lc07} introduced
two parameters ($\lambda_{TO}$ and $\alpha_{ex}$) to characterize
the transition disks. $\lambda_{TO}$ is
the wavelength at which the infrared excess begins and $\alpha_{ex}$
represents the SED slope at wavelengths longer than $\lambda_{TO}$. They showed
a diversity in $\alpha_{ex}$ of WTTSs and interpreted it as a wide
range of inner disk radii and temperatures compared to classical T Tauri
stars (CTTSs). We have also found several
stars in NGC 2264 whose SED indicated the existence of transition disks. 
Some stars showed a change of SED slope at [5.8], but many did so at [8.0].
In general the difference in $\lambda F_\lambda$ between [5.8] and [8.0] was
not very large. For simplicity and to avoid ambiguity due to errors in the
SED slope we have assumed that $\lambda_{TO}$ occurred at [8.0].
If $\lambda F_\lambda$ at [8.0] were smaller than the quantity at [24]
and the sign of the SED slope $\alpha$ from the IRAC bands ($\alpha_{\rm IRAC}$)
and that between [8.0] and [24] ($\alpha_{\rm LW}$) differed, then we classified
the object as a Class II object with a pre-transition disk if
$\alpha_{\rm IRAC} $ = -0.3 -- -1.8
or as an object with a transition disk if $\alpha_{\rm IRAC} <$ -1.8. 
In some cases, $\lambda_{TO}$ occurred at [5.8]; for these we calculated and 
used $\alpha_{\rm IRAC}$ between [3.6]
and [5.8] rather than $\alpha_{\rm IRAC}$ for all IRAC bands.
The distribution of objects with pre-transition or
transition disks in the C-C diagrams are shown in Figure \ref{figtdisk}.
In the ([3.6]-[4.5], [5.8]-[8.0]) diagram, objects with pre-transition
disks are located in the Class II locus, but most of the objects with
transition disks are found in the Class III/Photosphere locus or around
that. But in the ([3.6]-[4.5], [8.0]-[24]) diagram, such classes show strong
24 $\mu m$ excess emission and often fall on the border between Class I and II.
As already noticed by \citet{cjl06}, these are the objects with the largest
contrast between inner and outer disk emission and possibly the disks
with inner holes. The objects with pre-transition disks may be YSOs with
two disks with a gap between the inner and outer disks (see Fig. 4 of
\citet{ce08}).

\placefigure{figtdisk}

Interestingly, more than half (7 out of 13) of the objects with pre-transition
disks are in the Spokes cluster - the youngest region (see \S 4.2) in NGC 2264
- and therefore the pre-transition disks are occurring in an early stage of 
disk evolution. In addition, 18 out of 24 objects with transition
disks are in the halo region of the Cone nebula (Cone (H) in \S 4.1) and the
Halo region. The fraction is about 29\% relative to Class I and II objects 
in the Cone (H) region and about 14\% in the Halo region. Furthermore,
these objects (see  Figure \ref{figopt}) are
strongly concentrated between $I_C$ = 13 -- 16 in the PMS locus, corresponding
to masses about 1 solar mass (see \S 5 for details) at an age of
about 3 Myr \citep{sbc04}. This suggests that the inner disks of at least
some solar mass stars are evaporating rapidly and an inner hole is being
created at this epoch.

We also classified objects with PAH emission features.
When $\log \lambda F_\lambda$ at [8.0] exceeded the line drawn between
[5.8] and [24] by at least 0.2 dex and the difference was greater than 3 times
the error in $\lambda F_\lambda$ at [8.0], we classified the object as having 
a PAH feature. A total of 9 such objects were identified. 
One of them, W90 (= S2144 = SST9597) is a well-known Herbig Be star.
S2975 (= SST12833) also showed emission in H$\alpha$ and must also be
a Herbig Ae star.   Our H$\alpha$ photometry suggests that there is no
detected H$\alpha$ emission for
SST10184 (= C30920+C30962, C30920 dominates the light).
This new BMS star is about 0.5 mag fainter than the lower envelope of the 
PMS locus and is a Class II object.
Although two objects (SST 11099, 10914) have no optical counterparts, they
are probably embedded massive YSOs as they are in active star forming regions
(SST 11099 is in the Spokes cluster and SST 10914 in the Cone (C) region).
We could find no optical counterparts for the other four objects.
They are relatively faint ([3.6] $>$ 15 mag) and
probably are SFGs with PAH emission.

\begin{figure}
\epsscale{1.0}
\plotone{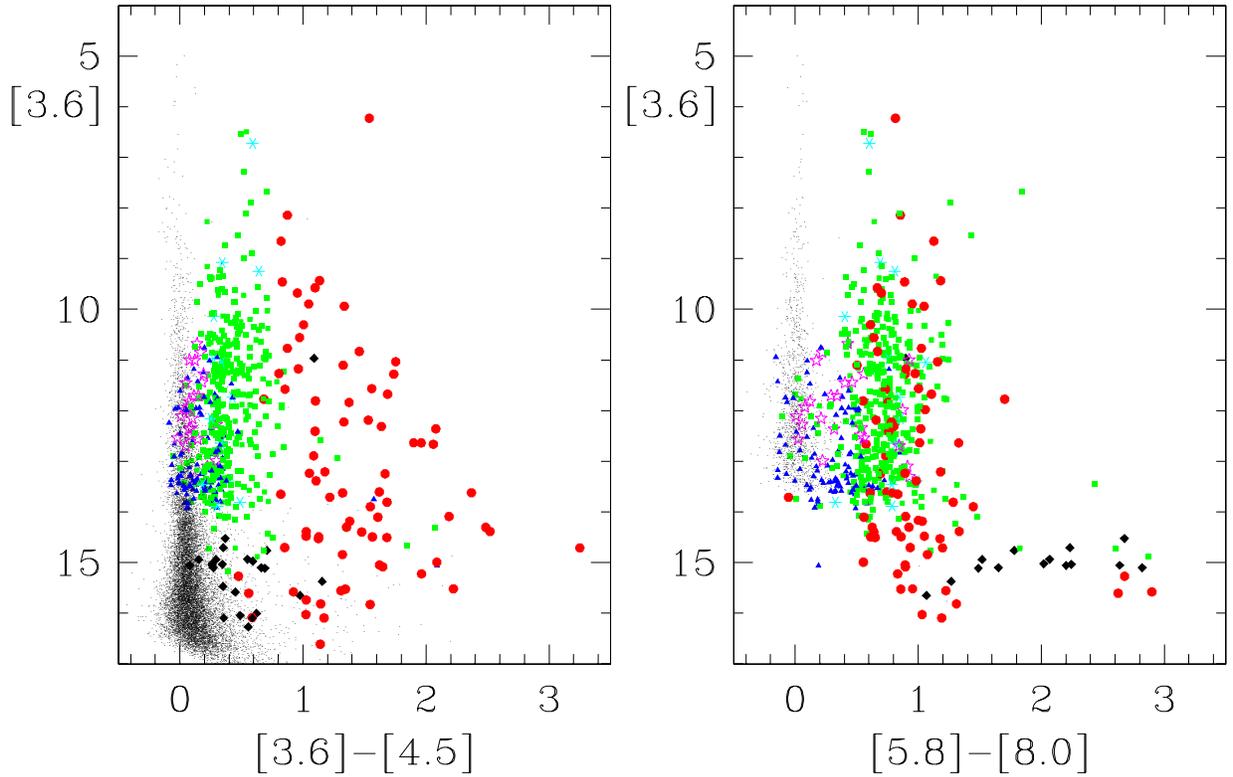}
\caption{The color-magnitude diagrams. Large (red) dot, (green) square, and 
(blue) triangle represents Class I, II, and II/III, respectively. (Black) diamonds
in the lower part are visually confirmed galaxies. Small dots are objects
that do not show any evident excess emission in {\it Spitzer}/IRAC passbands,
while (magenta) star symbols and (cyan) asterisks represent objects 
with transition disks and pre-transition disks, respectively.
\label{figcmd} }
\end{figure}

\begin{figure}
\epsscale{0.5}
\plotone{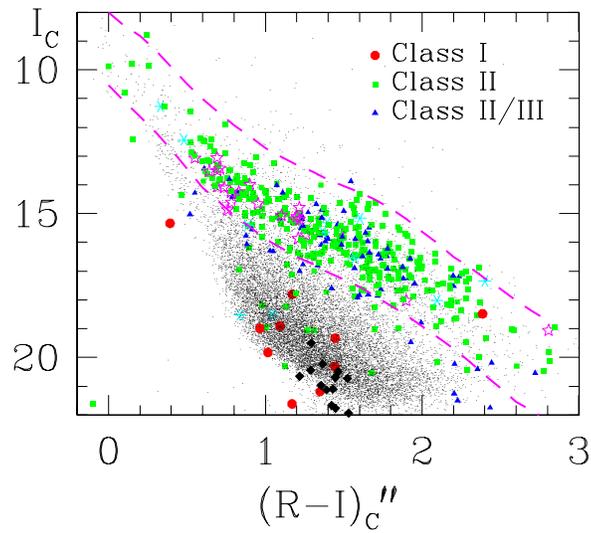}
\caption{The $I_C$ versus ($R-I$)$_C''$ color-magnitude diagram of stars
detected in the {\it Spitzer}/IRAC photometry. The dashed lines represent the locus
of PMS stars in NGC 2264 \citep{sbcki08}. Other symbols are the same as 
in Figure
\ref{figcmd}. Most Class I objects except one (C35527) are fainter and bluer
as are external galaxies. In general, most Class II objects are well
located along the PMS locus of NGC 2264 (see \citet{sbcki08}) but there are a
non-negligible number of Class II objects found below the PMS locus.
\label{figopt} }
\end{figure}

\begin{figure}
\epsscale{1.0}
\plotone{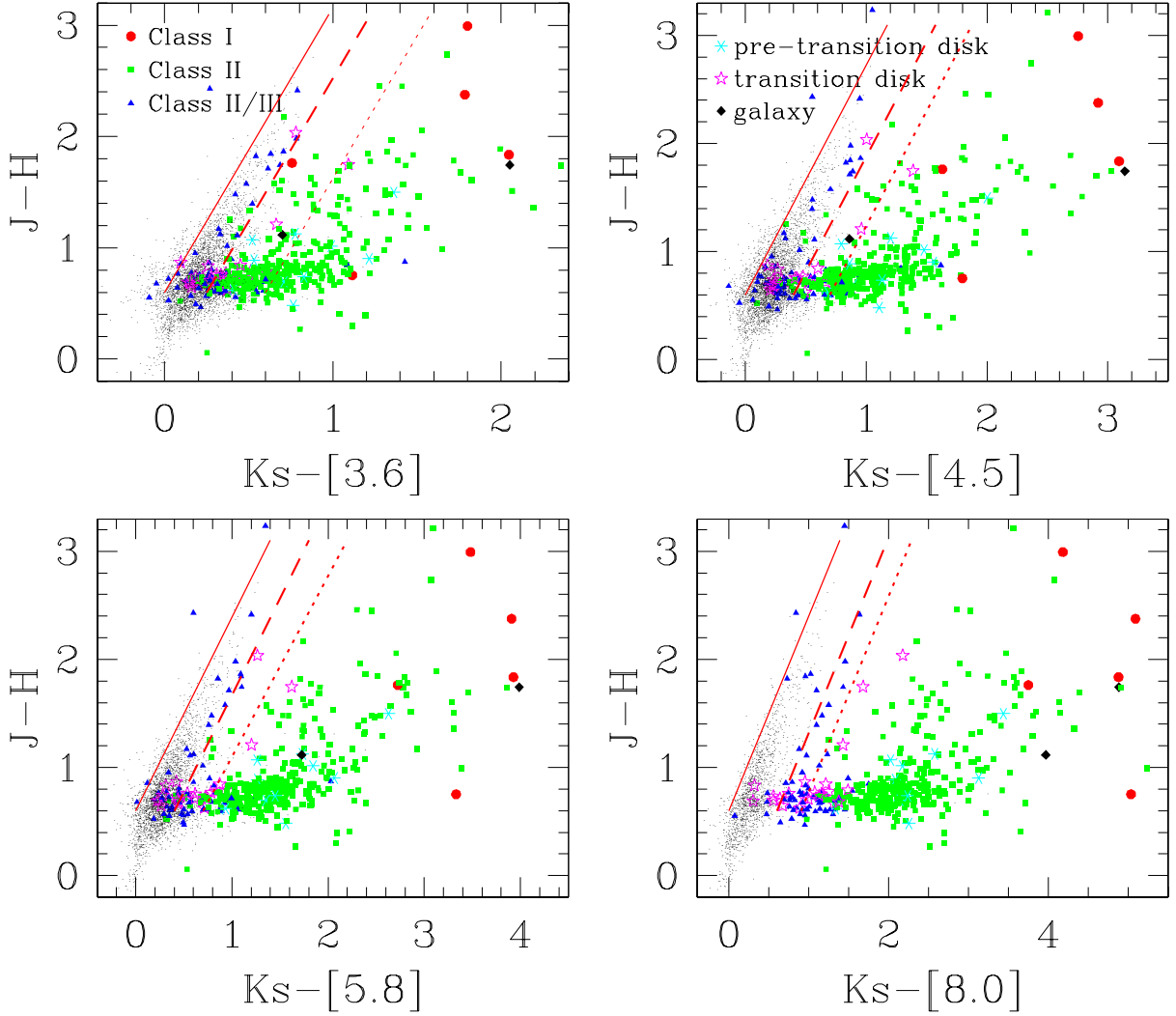}
\caption{The ($J-H$, $K_S$-[IRAC]) color-color diagram.
The solid and dashed lines in the diagrams represent, respectively, 
the apparent blue and red ridge of normal (reddened or unreddened) stars.
The region between the dashed and dotted line denotes the locus of (reddened
or unreddened) late-M type stars. The meaning of the symbols is explained in the
upper two panels. \label{fig2ms} }
\end{figure}

We present two CMDs in Figure \ref{figcmd}.  In this figure, and the next
two figures, the YSO classes used are those from our $Q_{CC}$ scheme.
In the [3.6] versus [3.6]-[4.5] diagram, Class I (big red dot) and Class II
(green square) objects are well separated with few exceptions, but in the
[3.6] versus [5.8]-[8.0] diagram, the separation is not evident. The [3.6]-[4.5]
color of objects with (pre-)transition disks (see \S 3 for the classification
scheme) is close to that of normal objects without disks. But they are redder
in [5.8]-[8.0] than normal stars. In the
second CMD, we can see many external galaxies (black diamonds) in the faint,
red portion of the diagram (identified as galaxies via
examination of our deep CFHT images). There is a very bright galaxy (SST 6262 = C22243
= 2MASX J06402351+0956312; [3.6] $\sim$ 11 mag). It is apparently
an edge-on spiral galaxy (Type in NED: IrS).

\placefigure{figcmd}

We present an $I_C$ versus ($R-I$)$_C''$ diagram\footnote{($R-I$)$_C''$ is
($R-I$)$_C$ corrected for the effect caused by use of the Mould $R$ filter
mentioned in \citet{sbcki08}.} in Figure \ref{figopt}. Most Class II objects
are located along the PMS locus. Some Class II objects are located below
the PMS locus and are BMS stars. On the other hand, the few Class I objects 
with optical photometry are, on average, relatively fainter and bluer
than Class II objects.   The extragalactic objects also tend to be very
faint and thus lie below the PMS locus.
Interestingly, many objects with transition disks (star
symbols in Figure \ref{figopt}) are concentrated between $I_C$ = 13 -- 16.

\placefigure{figopt}

We present the combined 2MASS-IRAC C-C diagrams in Figure \ref{fig2ms}.
The solid and dashed lines represent the locus of reddened or unreddened
normal stars. The region between the dashed and dotted lines is the locus
of (reddened or unreddened) late-M type stars. Disked stars are well separated
from normal stars at the longer wavelengths, but many of them do not show
an appreciable IR excess in the ($J-H$, $K_S$-[3.6]) diagram and are located
in the region occupied by late-M type stars. One Class I object C35527 does not
show any excess emission in this diagram. That may be due to variability
of the object (see Figure \ref{figsed} and \S 6.3.2). There is no clear
separation between Class I and II objects in the figures, at least in terms
of $J-H$ excess. This fact implies that
the size of the near-IR excess is not strongly related to the evolutionary
stage of YSOs.   Few Class I objects appear in these diagrams because most
of the Class I sources are too faint to be detected at $J$ by 2MASS.

\placefigure{fig2ms}

\subsection{Final Classification and Comparison}

\begin{figure}
\epsscale{0.5}
\plotone{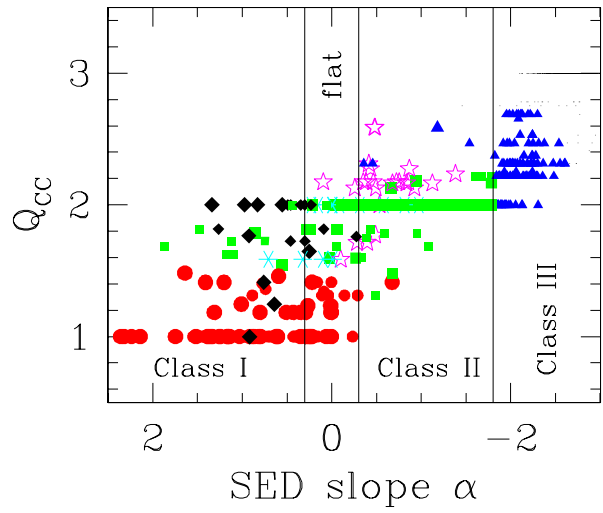}
\caption{Comparison of SED slope and photometric criterion $Q_{CC}$.
Symbols are the same as in Figure \ref{figcmd}. In general the two criteria
correlate well.  \label{figcls} }
\end{figure}

\placefigure{figcls}

\begin{figure}
\epsscale{1.0}
\plotone{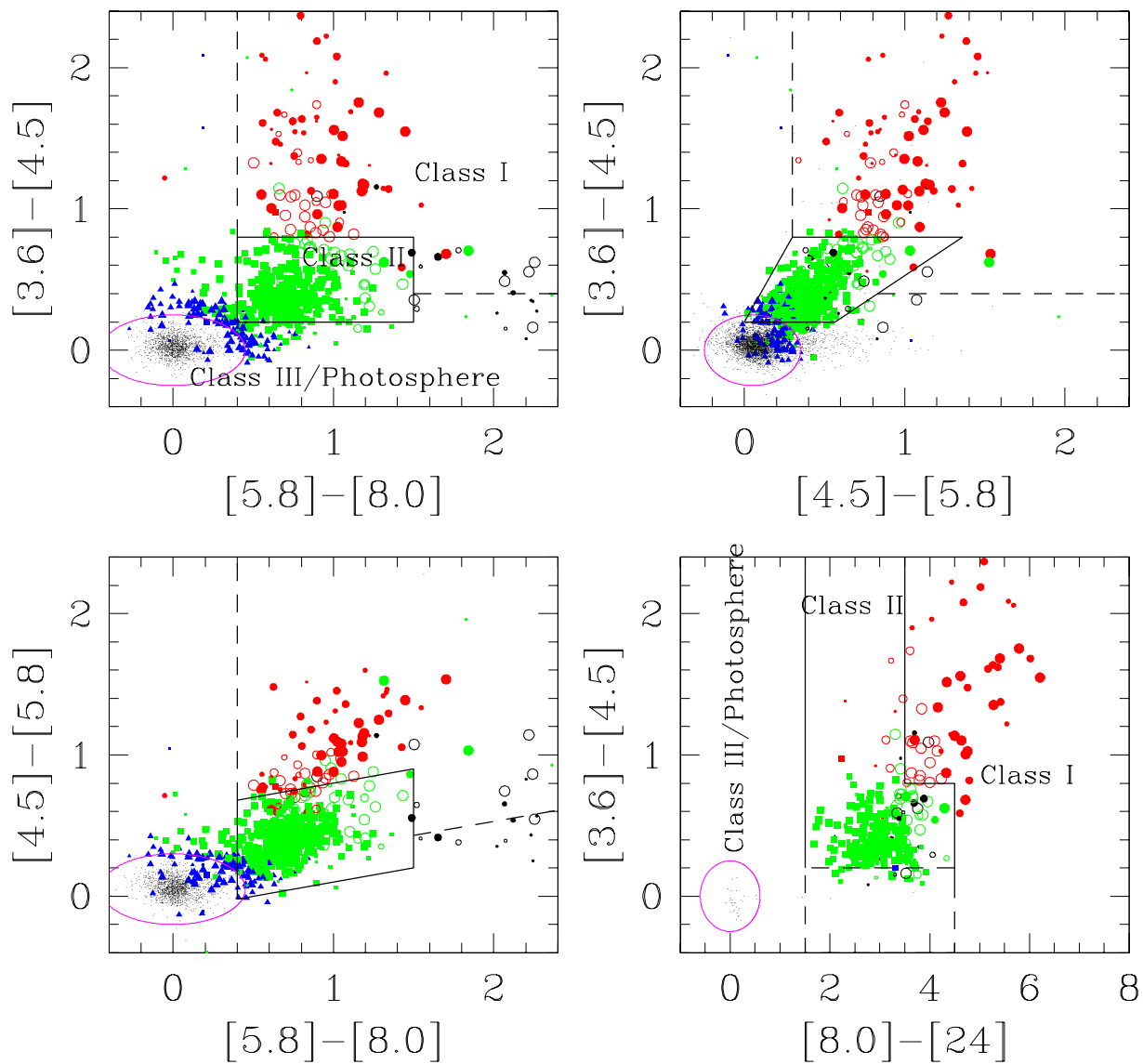}
\caption{Comparison of SED slopes and photometric classification criteria.
The color represents the class from photometric criteria, while the symbols
denote the class from the SED slope. The size of the symbols are inversely
proportional to
the error in the SED slope. The color of symbols are the same as
in Figure \ref{figcmd}. Filled circles, open circles, squares, and triangles
represent respectively Class I, flat, Class II, and Class III objects.
\label{figcmp} }
\end{figure}

The photometric classification criteria $Q_{CC}$ are compared
with the SED slope $\alpha$ in Figure \ref{figcls}. In general, $Q_{CC}$ is
well correlated with the slope $\alpha$. There are several exceptions.
One is W90 (= S2144 = SST 9597). As can be seen in Figure \ref{figsed},
the slope $\alpha$ of W90 is +0.56, suggesting the star
is a Class I object. But originally a Class I object was a YSO in the early
evolutionary stage, i.e. a YSO with an accretion envelope. In this respect
W90 is not a Class I object. We kept the photometric classification for W90.
If the two criteria conflicted, in general we adopted the later evolutionary
stage. In many cases such conflicts were caused by errors in the SED 
slope. Some of the objects have properties that 
are very close to the border between Class I and II in
Figure \ref{figcmp} - their actual classification is therefore somewhat
ambiguous.

\placefigure{figcmp}

We also compare the results from the two classification schemes in Figure
\ref{figcmp}. The symbol-type represents the Class from the SED slope, while
the color denotes the class from the photometric criteria. The size of
the symbols is inversely proportional to the error in $\alpha$. 
We did not mark objects with pre-transition or transition disks. 
As can be seen in Figure \ref{figcmp}, the two results match quite well. 
The objects with flat spectra (open circles) are scattered along the border 
line between Class I and II. Photometric Class II objects with flat spectra 
are distributed in the red part of the Class II locus. Class II/III 
objects are distributed around the outer boundary of the Class III/Photosphere
objects. The YSO class from our final classification is listed in the 20th
column of Table \ref{tab_sst}.
 
An aim of this study was to obtain the fraction of stars with disks.
To do that, we need to identify a subset of the NGC 2264 population where
we believe we have a nearly complete list of members.   We believe that
the best choice for this is
the S MON and CONE regions of \citet{sbcki08}, where our membership
completeness
was very high. We only consider stars in the PMS locus ($I_C$ = 8 -- 15.5).
$I_C$ = 15.5 was set from the 80\% completeness of all IRAC
channels as mentioned in \S 2.3. Among 137 known CTTSs, 76 were 
Class I or II (objects with primordial disks) and 22 were Class II/III 
(objects with optically thin disks). The disk detection fraction for CTTSs
was 71.5\%; the primordial disk fraction (counting only the Class I or
II objects, was 55.5\%). 
There are 181 objects in the region originally classified as WTTSs based 
on their being believed to be members but having weak or undetected H$\alpha$
emission. Of these 181 classified as WTTSs based on H$\alpha$, we find
14 objects are Class II based on their IR excesses and 10 are Class II/III
objects. The disk detection fraction
of WTTSs was 13\% and the fraction of objects with primordial disks
was only about 8\%. Two objects were newly identified PMS stars in NGC 2264
from {\it Spitzer} observation alone.   
\citet{lc07} obtained about 20\% for the
primordial disk fraction of WTTSs in Ophiuchus, Lupus and IC 348, while
\citet{cjl06} found that about 12\% of the optically classified WTTSs in
IC 348 had primordial disks. On the other hand, \citet{dlp06}
obtained a much lower disk fraction of about 6\% for relatively isolated WTTSs.
Our value for NGC 2264 is between these extremes.
For all PMS objects (CTTS plus WTTS), our disk detection fraction is 38\%.

\section{DISTRIBUTIONS OF YOUNG STELLAR OBJECTS}

\subsection{Spatial Distributions}

\begin{figure}
\epsscale{1.0}
\plotone{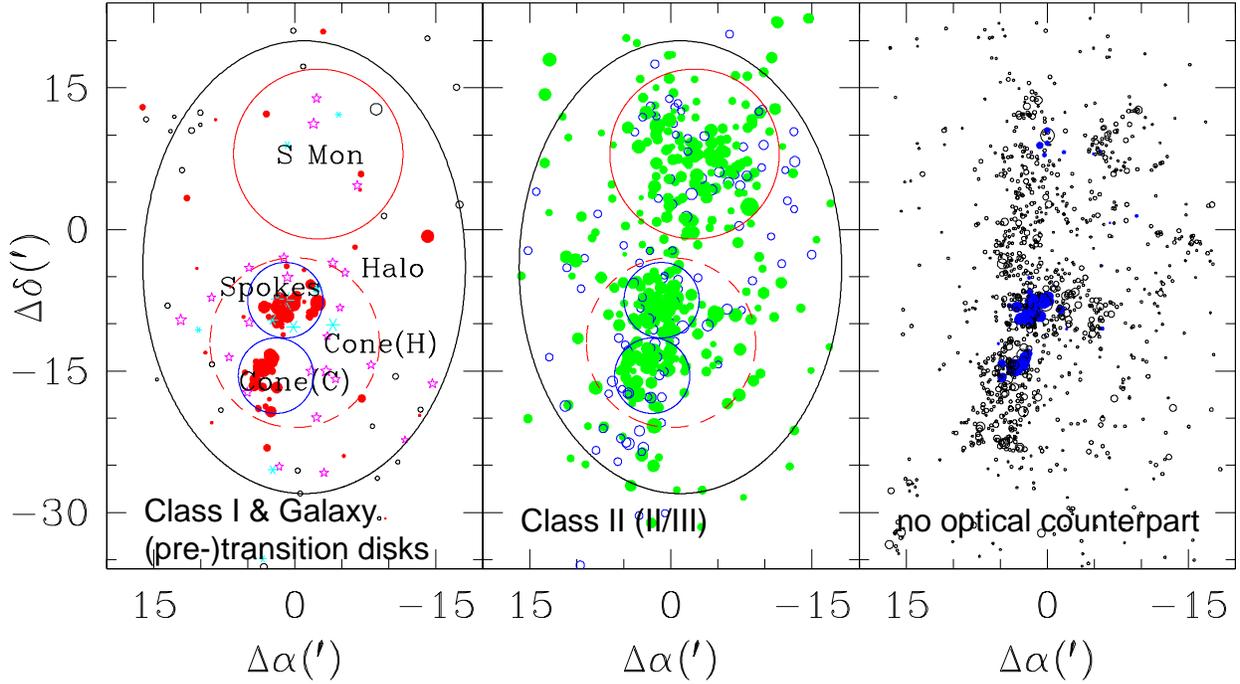}
\caption{(left) The spatial distribution of Class I objects (red dots) and
optically confirmed galaxies (open circles). Asterisks and stars represent,
respectively, objects with pre-transition disks and with transition disks.
(middle) The distribution of
Class II (green dots) and II/III objects (blue circles).
(right) The distribution of optically invisible IR sources. Blue dots
represents X-ray emission objects with no optical counterpart.
\label{figmap} }
\end{figure}

The spatial distribution of YSOs contains information about both where
they are formed and what happens after that due to dynamical evolution.
Based on the density distribution of H$\alpha$ emission stars, two active
star forming regions (S MON and CONE) and the HALO region surrounding these
two subclusterings were identified in \citet{sbcki08}. Mid-IR images obtained
with the {\it Spitzer} Space Telescope gives invaluable information on
the embedded YSO population. The spatial distribution of YSOs in NGC 2264
is shown in Figure \ref{figmap}. Most Class I objects are concentrated
in either the Spokes cluster 
($\Delta \alpha$ = 1.$'$0, $\Delta \delta$ = -7.$'$5)
or the Cone nebula region ($\Delta \alpha$ = 2.$'$0, $\Delta \delta$ = -15.$'$5
- Cone (C)). These two subclusterings belong to the CONE region as defined
in \citet{sbcki08}.
The Class II objects are more widely distributed, but the distribution of Class II
objects definitely resembles the distribution of H$\alpha$ emission stars
in \citet{sbcki08}. Now, based on the spatial distribution of YSOs detected from
{\it Spitzer} observation, the definition of the SFRs should be refined.
The CONE nebula region in \citet{sbcki08} should be divided into three regions
- the Spokes cluster \citep{pst06}, the Cone (C) (C means ``core'') and Cone (H)
(H means ``halo'') regions. The spatial distribution of pre-transition disks and
transition disks is shown in the left panel of Figure \ref{figmap}.
The other regions are the same as in \citet{sbcki08}.
The one bright Class I object in the HALO region is C13507 (= SST 2558 = R04-45; 
see \S 2.5). \citet{rpabv04} classified the object as being of 
H$\alpha$ emission class 4.5 (prominent in H$\alpha$).

\placefigure{figmap}

In the right panel of Figure \ref{figmap}, we plot the spatial distribution
of IR objects with no optical counterpart. Blue dots represents X-ray emission
objects (see \S 2.3). Interestingly, most X-ray sources without
optical counterparts are coincident with the location of the two embedded clusters.
This fact favors our contention that these objects are bona fide members of the two embedded
subclusterings. The other X-ray emission sources surround the most massive
star S Mon. The optical counterpart of these objects may not be detected
in deep optical images due to the scattered light from S Mon - the optically
brightest star in NGC 2264 (eg. PID 165 in Table 3 of \citet{sbc04}).

\subsection{Frequency Distributions of SED Slopes}

\begin{figure}
\epsscale{1.0}
\plotone{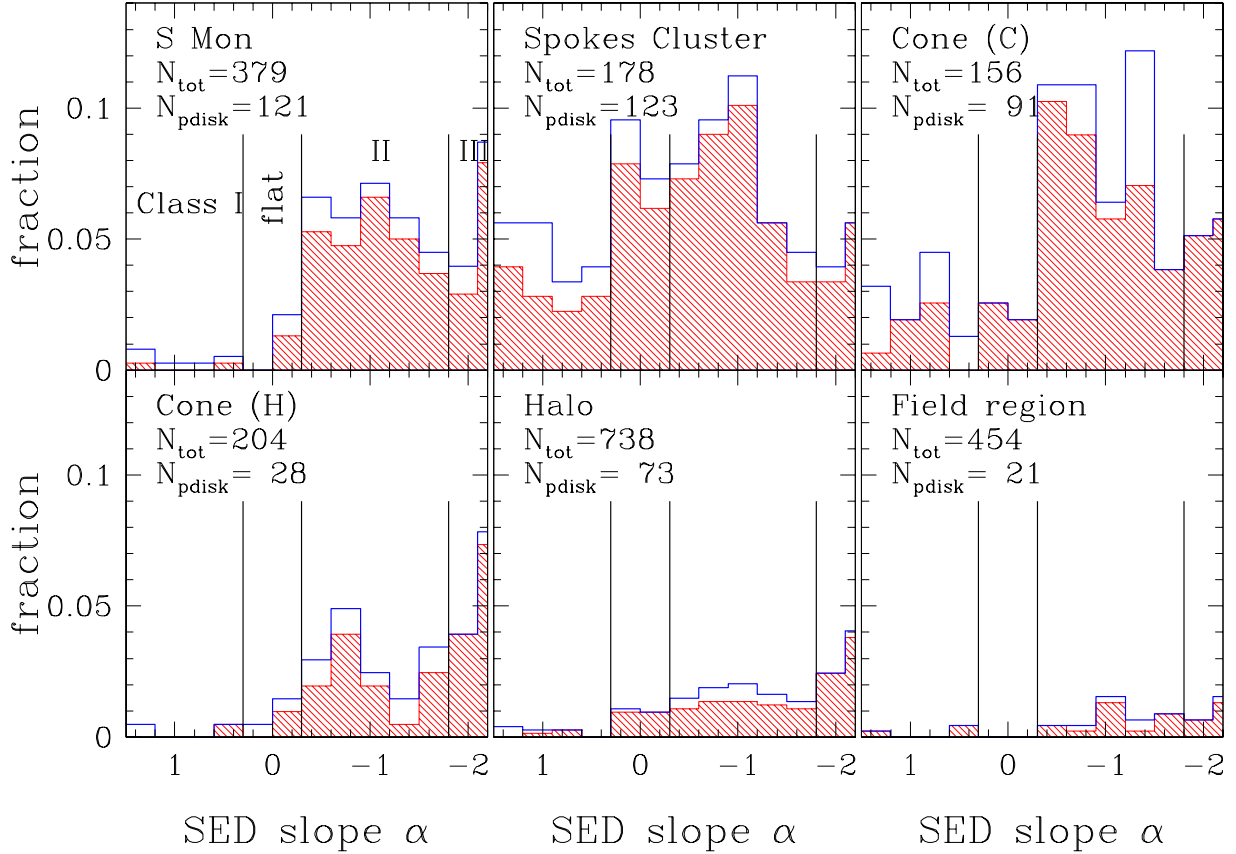}
\caption{The distributions of SED slope $\alpha$ for 
all objects in a given region.
The hatched histogram represents the fraction of $\alpha$ with a small error,
while  the white histogram denotes the fraction of $\alpha$ regardless of error.
N$_{\rm tot}$ represents the 
total number of objects having good data in the region,
while N$_{\rm pdisk}$ means the number of objects having $\alpha \geq$ -1.8.
For the S Mon region, Class II objects are dominant. The fraction of Class I
objects is high in the Spokes cluster. Cone (C) is intermediate between the 
Spokes cluster and S Mon. The other three regions are dominated by Class III/
Photosphere objects. \label{figdist1} }
\end{figure}

\begin{figure}
\epsscale{1.0}
\plotone{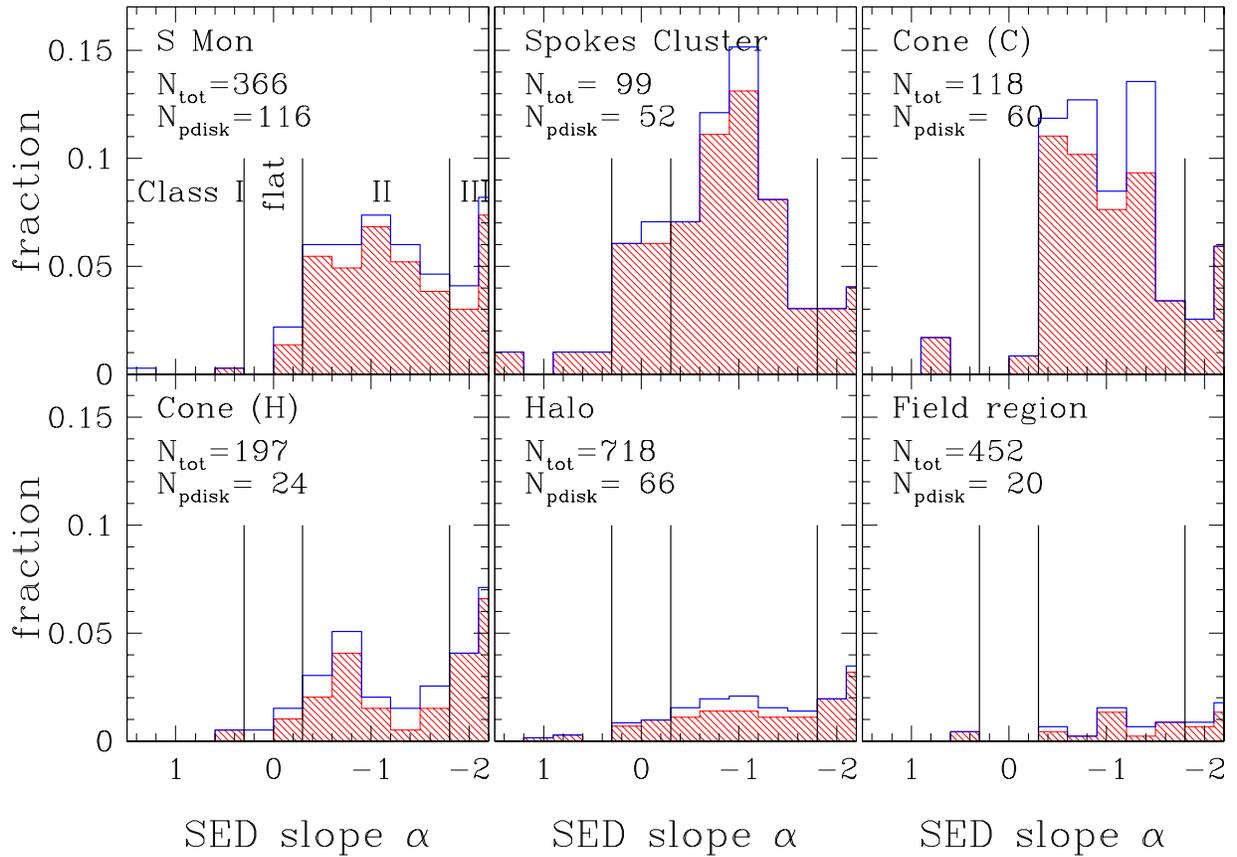}
\caption{The distributions of slope $\alpha$ of optically visible objects.
Symbols are the same as Figure \ref{figdist1}.  \label{figdist2} }
\end{figure}

The distribution of SED slopes for a given SFR can give information on
the evolutionary stage of the SFR. Figure \ref{figdist1} shows the distribution
of all objects (detected in at least four IRAC bands and where the total error in the
four IRAC bands is less than 0.25 mag) in a given SFR, while Figure
\ref{figdist2} represents that of optically visible objects in the SFRs.
The hatched histogram represents the distribution of $\alpha$ with smaller
error ($\epsilon (\alpha) \leq 0.3$), while the open histogram denotes
the distribution of all samples.

\placefigure{figdist1}
\placefigure{figdist2}

The distribution of $\alpha$ is far different in the upper three panels.
Class II objects are dominant in the S Mon region, while the fraction of Class I
and flat spectrum objects is very high in the Spokes cluster. 
Cone (C) is intermediate between these two regions. Most of the Class I objects 
are invisible down to $I_C$ =22 mag, 
i.e. embedded objects. These facts imply that the Spokes cluster is
the youngest and S Mon is the oldest. The average values of $\alpha$ for
$\alpha > -2.5$ are -1.31, -0.51, -0.80 for S Mon, the Spokes cluster and 
the Cone (C) region, respectively. 
(If the whole range of $\alpha$ is considered, the average values
of $\alpha$ are -2.01, -0.84, -1.67 for S Mon, the Spokes cluster and the 
Cone (C) region, respectively.) 
If we accept the gradual evolution of SEDs, the difference
in the mean value of $\alpha$ among subclusterings also demonstrates the
sequence of star formation within NGC 2264. Although there are several
embedded Class II objects
in the Spokes cluster, the physical connection between the embedded
and exposed population (in Figure \ref{figdist2}) is unclear. The continuous distribution of $\alpha$
and the spatial concentration of both Class I and II objects in the 
Spokes cluster
may imply a physical connection between the embedded population and
the exposed population. In addition, recently \citet{lmr07} found several young
stellar aggregates with a wide range of SEDs in the Perseus SFR.

The other three regions in the lower panel are 
dominated by Class III/Photosphere objects.
Interestingly, the peak of the distribution occurs at $\alpha$ = -2.75 $\pm$ (Halo: 0.19,
Field: 0.16). This value is very close to the expected value of $\alpha$
($\alpha \approx$ -2.8 for an average over stellar spectral types --
\citet{ph06}). In NGC 2362 ($A_V$ = 0.3 mag), the peak distribution of stars
without disks occurs at $\alpha$ = -2.85 $\pm$ 0.17 \citep{dh07}. In IC 348
and Taurus SFR, the peak frequency of $\alpha$ occurs at about -2.7
\citep{cjl06}. The slight difference in $\alpha$ of the peak distribution
among young open clusters could be interpreted as a difference in reddening
as the SED slope 
is affected by reddening\footnote{The amount of reddening can alter
the slope by $ {{d \alpha} / {d E(B-V)} }\propto$ 0.0224 mag$^{-1}$ for
the reddening law adopted in \S 3.2. If we assume that the difference in 
$\alpha$ ($\equiv \alpha_{peak} - \alpha_{expected} \approx$ -0.05) is
solely caused by reddening, the derived reddening is about $E(B-V) \approx$
2.2 mag.}.

\section{DISK FRACTION}

The frequency and lifetime of disks around young stars are important topics
of recent research because planets are believed to form in the disks.
The lifetime of a disk should be at least longer than the time required for
planet formation. In addition, the frequency of long-lived disks should be
consistent with the frequency of exoplanets in the solar neighborhood.
From a survey of several young open clusters in the $L$-band, \citet{hll01}
concluded that the initial disk fraction is as high as 80\% and that the disk
fraction decreases rapidly with a half-lifetime of about 3 Myr. They found
that about 50\% ($\pm$ 10\%) of stars in NGC 2264 showed a $JHKL$ excess.
\citet{cjl06} estimated the total frequency of disk-bearing stars in the
2 -- 3 Myr-old young open cluster IC 348 to be 50\% $\pm$ 6\% and that of
stars surrounded by primordial disks to be 30\% $\pm$ 4\%.
Recently, \citet{dh07} determined a low value ($\sim$ 7\% $\pm$ 2\%) for
the primordial disk fraction in the 5 Myr-old young open cluster NGC 2362.
As the age of NGC 2264 is about 3 Myr \citep{sbc04}, information
on the disk fraction will be valuable data for the disk evolution and 
the lifetime of primordial disks around young stars.

\begin{figure}
\epsscale{0.5}
\plotone{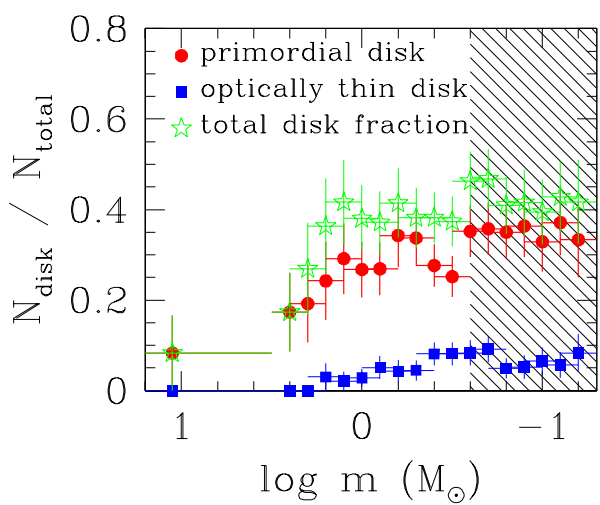}
\caption{Disk fraction of stars in NGC 2264. The mass of PMS stars
are estimated using the PMS evolution models of \citet{sdf00} and 
\citet{bcah98} for very low-mass stars. (Red) dots, and (blue) squares represent
the minimum fraction of stars with primordial disks and with optically thin
disks, respectively. The bar represents the error from Poisson statistics.
The stars represent the total disk fraction, i.e. the sum of primordial disks,
optically thin disks, pre-transition disks, and transition disks. The hatched
portion represents the range where membership selection is incomplete.
\label{figdiskf} }
\end{figure}

\placefigure{figdiskf}

We present the disk fraction of stars in the PMS locus of NGC 2264 
as a function of stellar mass in Figure \ref{figdiskf}. As the membership
selection is fairly incomplete in the HALO and FIELD regions of 
\citet{sbcki08}, we consider only the member stars in the S MON and 
CONE (Spokes + Cone (C) + Cone (H)) regions. The mass of PMS members in
the PMS locus of NGC 2264 (see Figure \ref{figopt}) is derived using
the PMS evolution model of \citet{sdf00} for mass $>$ 0.1 M$_\odot$ and 
using \citet{bcah98} for lower mass stars. Because the mass determinations
require optical photometry, the set of stars we considering excludes
the youngest, most embedded members, especially in Spokes and Cone (C).
The theoretical parameters are transformed to the observational CMD using
the empirical color-temperature relation for late-type stars of \citet{msb95}.
The distance
($V_0 - M_V$ = 9.4 mag) and reddening ($E(B-V)$ = 0.07) are from \citet{sbl97}.
The fraction of primordial disks is the ratio of the number
of Class I and Class II stars to total member stars for a given mass range.
The fraction of optically thin disks is that of Class II/III stars to the total
number of member stars. For clarity in the figure, we did not mark
the fraction of pre-transition and transition disks. But the total disk
fraction is the ratio of all disked stars (primordial disk + optically thin
disk + pre-transition disk + transition disk) to total member stars for
a given mass range.
The disk fraction presented in Figure \ref{figdiskf} was derived in
logarithmic mass intervals of $\Delta \log m$ = 0.2 for $\log m <$ 0.5,
and then recalculated using the same bin size but shifted 
by 0.1 in $\log m$ to smooth
out the binning effect. The hatched area represents
the region where membership selection is somewhat incomplete.

The fraction of stars with an primordial disk is about 28.5\%
for $\log m$ = 0.2 -- -0.5 and slightly higher (about 35\%) for $\log m$ = 
-0.2 -- -0.3.  The fraction of optically thin disks is generally very low,
but slightly higher for low-mass stars. The total disk fraction is, on the
other hand, nearly constant (about 38\%) between $\log m$ = 0.3 -- -0.6.
Interestingly, the fraction of transition disks is very high in the range
 $\log m$ =0.1 -- -0.1 (about 8.5\%). 
This fact suggests that at the age of the non-embedded population of NGC 2264 (about
3 Myr - \citet{sbc04}) disk evaporation, especially the inner disk, of solar
mass stars is very active. If we include the fraction of 
pre-transition disks and transition disks to the fraction of primordial disks,
and consider them all as a fraction of the primordial disks, then the disk fraction
of solar mass stars is slightly higher than that of lower-mass stars.

\section{EFFECT OF MASSIVE STARS on DISKS}

\subsection{Disk Fraction of Young Stars around S Monocerotis}

The idea that the formation of massive stars 
could influence the fate of natal clouds as well as its low-mass siblings 
can be traced back to the paper by \citet{ghh62}.
{\it Hubble} Space Telescope images of the Orion nebula
have revealed clearly the effect of the massive star $\theta^1$C Ori (O6pec) on
surrounding low-mass young stars, especially the proto-planetary disks around
them \citep{odw94}. Recently \citet{sbcki08} showed that the strength of
H$\alpha$ emission $\Delta (R_C'-H\alpha)$ decreases systematically with
increasing distance from the massive star S Mon (O7V) or HD 47887 (B1.5Vp). 
In addition, the $^{13}$CO map of NGC 2264 \citep{gf06} shows
a bubble-like feature at the position of the most massive star S Mon.

\begin{figure}
\epsscale{0.5}
\plotone{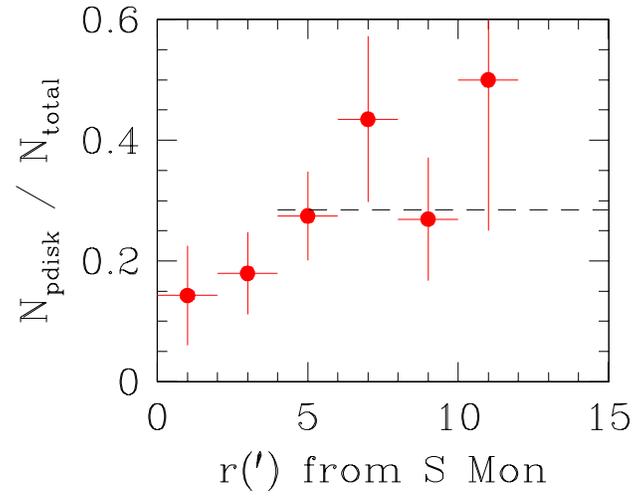}
\caption{The radial variation of the primordial disk fraction from S Mon.
The thin lines represent the errors from Poisson statistics. The meaning of the
other symbols is the same as Figure \ref{figdiskf}.
\label{figdvar} }
\end{figure}

\placefigure{figdvar}

It is important to look for clues relating to the effect of massive
stars on the fate of the disks around low-mass stars. We calculated
the variation of disk fraction as a function of distance from the most
massive star, S Mon; the result is shown in Figure \ref{figdvar}. In the calculation, we used
data for stars with masses between $\log m$ = 0.3 -- -0.6 to minimize
effects due to incompleteness in the data. Interestingly, the fraction of
primordial disks increases as the distance from S Mon increases and approaches
the average value of primordial disk fraction at $r \approx$ 6$'$. [N$_{pdisk}$ 
/ N$_{total}$ $\approx$ 0.100 ($\pm$ 0.030) + 0.033 ($\pm$ 0.009) $r'$, 
r = 0.97 for $r \leq 5.'6$). In addition, there are no disked star within 
$r = 1'$. These observations suggest strongly that
the ultraviolet photons and stellar wind from the massive O7V star S Mon 
has actually affected
the fate of the primordial disks around young low-mass stars up to $d \approx$
1.2 pc. Recently \citet{zb07} fond a similar trend in NGC 2244.
We also checked for variation in disk fraction with distance from the 
second most massive star, HD 47887 (B1.5Vp), but found nothing.

\subsection{Disk Fraction Versus Age Relation}

Prior to Spitzer's launch, \citet{hll01} performed an $L$-band
survey of young open clusters and
derived the half-lifetime for disks around young stars.   Because the
disk lifetime
is a function of stellar mass \citep{cjl06,dh07}, the disk fraction versus
age relation could be affected by the limiting magnitude of the survey.
Now the primordial disk fractions based on {\it
Spitzer} observation has been calculated for more than a dozen of young open
clusters and associations.   The   {\it Spitzer} version of the ``Haisch''
diagram has already been determined (for example, see \citet{hj07a}).
However, There are several reasons for reexamining the relation. In many
cases
the mass range was not taken into account; in addition, membership
selection seems to be incomplete because many researchers calculated the disk
fraction based on the members selected from {\it Spitzer} observations only.
In that case many young stars without detectable excess emission in the
mid-IR may have escaped membership selection.

\subsubsection{NGC 1333 \& Serpens}

\citet{rag08} presented the background-subtracted $K_s$ luminosity function
for the young embedded cluster NGC 1333. The fraction of IR excess YSOs of
bright $K_s$ objects is slightly higher that that of fainter low-mass
objects. The fraction of IR excess stars for $K_s \leq 12.5$ is about 80 ($\pm$ 16) \%,
while that of low-mass stars ($K_S$ = 10.5 -- 12.5, equivalently m = 0.2 --
1.0 M$_\odot$) is about 73 ($\pm$ 16) \%. Such a difference in the objects with
disks could be caused by the problem in statistical subtraction due possibly to spatially
varying reddening, and therefore we adopt 80\% as the primordial disk fraction
of NGC 1333. 
Another young embedded star forming core Serpens was studied by \citet{we07}
with the data from {\it Spitzer} space telescope and {\it Chandra} X-ray observatory (CXO).
They used the criterion of Class II objects as SED slope $\alpha$ between
-2.0 and -0.3. For the completeness of membership selection, we limit
our analysis to objects within the {\it Chandra} FOV. In addition for the consistency
of classification used in this paper, we exclude 6 objects with $\alpha <$
-1.8 or a larger error in SED slope. In addition we only calculate 
the primordial disk fraction of YSOs brighter than $K_s <$ 12.6.
The fraction is only about 1 \% lower that that of all members in the {\it Chandra} FOV.

\subsubsection{NGC 2244}

The young open cluster NGC 2244 is one of the important targets for studying
the effect of massive stars on the evolution of disks around low-mass stars
because it contains about 30 early type stars between O5V and B3V.
\citet{pns02} performed optical photometry and determined the distance and
age.
Later, \citet{zb07} found 25 Class I and
337 Class II objects in the cluster based on {\it Spitzer} observation and
obtained a
slightly lower disk fraction of about 45\% for YSOs in NGC 2244.

Recently, \citet{wj08} studied the cluster using data obtained with
CXO. They found over 900 X-ray sources in the cluster.
We have re-analyzed the cluster using the optical images of \citet{pns02},
{\it Spitzer} data in \citet{zb07}, and X-ray sources in \citet{wj08}.
For completeness of membership selection we limit our attention to the
overlap region of the three observations. For X-ray observations, we limit
the FOV to that of CXO ObsID = 3750 because the exposure times of other
regions
was much shorter. There were 748 members in the region.
Among them 204 stars have masses between 1 M$_\odot$ and 0.5 M$_\odot$
(the completeness limit of X-ray selected members), with 74 stars
being classified as Class I, flat, and Class II.
The fraction of stars with primordial disks is about 36 ($\pm$ 5) \%.
This value is slightly lower than that obtained by \citet{zb07}.
The median age of the low-mass stars is 1.7 ($\pm$ 0.2) Myr.

\subsubsection{Ori OB1 association}

\citet{hj07a} studied the disk fraction of about 3 Myr-old open cluster
$\sigma$ Ori, and obtained a disk fraction of about 35 \% for T Tauri stars.
They listed 336 members, but many of them do not show variabilty or X-ray emission.
We selected 194 members based on X-ray emission, IR excess, and optical variability.
Among 194 members, 99 objects have masses between 1 M$_\odot$ and 0.2 M$_\odot$,
among these 99 low-mass stars, 48 are classified as Class II or YSOs with 
(pre-)transition disks. The primordial disk fraction of $\sigma$ Ori is about 48 \%.
There is no homogeneous optical data for the cluster and therefore it is
very difficult to determine the median age of the cluster. Various optical
data in their tables and from SIMBAD database were compiled, and the median
age was estimated to be 1.9 ($\pm$ 0.3) Myr from ($V,~ V-I$) diagram.

The disk fraction of 25 Ori region was studied by \citet{hj07b}. They listed
115 members. Among them 54 stars have masses between 1 M$_\odot$ and 0.2 M$_\odot$
(Sp = K7 -- M4 or $J$ = 12 -- 13.2). Only 1 star is classifed as Class II.
Normally the age of 25 Ori is assumed to be about 10 Myr, but Figure 4a
of \citet{bc07} implies a slightly younger age. We adopt the age of 25 Ori
as 8 (-1, +2) Myr. NGC 2068/2071 is another young SFR in Ori OB association.
\citet{fm08} presented {\it Spitzer} data, spectral type, and stellar paremeters
of 69 likely members. Among 69 likely members, 48 stars are low-mass
stars (Sp = K7 -- M4), and 26 out of 48 stars are classified as CTTS. And therefore
the primordial disk fraction of NGC 2068/2071 is about 54\%.

\subsubsection{IC 348 \& NGC 2264}

The young open cluster IC 348 in the Perseus SFR has been well studied by
\citet{cjl06}. They listed 307 members selected from X-ray emission,
IR excess, and spectroscopy. The number of low-mass members between K7
and M4 is 108, and 37 out of 108 stars
are classified as stars with thick disks. The primordial disk fraction
of low-mass stars is therefore about 34 ($\pm$ 6) \%. There are three
active SFRs in NGC 2264, but many of the YSOs in two of these SFRs (Spokes
cluster and Cone(C)) are an embedded population. The mass and age of these
YSOs cannot be determined reliably. We calculated the disk fraction of YSOs
in the S Mon region of NGC 2264. The completeness limit of membership
selection was set by the exposure time of the X-ray observations (see
\citet{sbc04,svr04}). The total number of low-mass members is 139, and
among them 37 stars were classified as stars with primordial disks.
The median age of low-mass stars in S Mon region of NGC 2264 is 3.1 Myr.

\subsubsection{$\gamma$ Vel}

One of the most important clusters in this study comprises the young stars
around WC star $\gamma$ Vel. \citet{hj08} found several debris disk
candidates among 579 candidate members based on the color-magnitude
diagrams. We have re-analyzed their photometric data and adopted a more
conservative set of membership criteria. The number of members we find
from X-ray emission and IR excess is only 141 stars. Among them 41 stars
are low-mass stars ($\log m =$ 0.0 -- -0.7) with only one Class II object.
The age of the cluster was assumed to be about 5 Myr \citep{hj08}, but
the median age of our selected members is 3.4 ($\pm$ 0.5) Myr. \citet{rdj09}
also obtained a similar age for $\gamma$ Vel. The primordial disk fraction
of $\gamma$ Vel is therefore about 2 ($\pm$ 2) \%, which is far lower
than any other similar age young SFRs.

\subsubsection{Cep OB2 association}

There are two young open clusters (Tr 37 and NGC 7160) in Cep OB2 association.
Tr 37 is younger, and its assumed age is about 4 Myr \citep{saa06} while that of
NGC 7160 is about 10 Myr. \citet{saa05,saa06} selected spectroscopic members
and studied the young stars in these clusters. There are 130 and 14 low-mass
members (Sp = K7 -- M4) in Tr 37 and NGC 7160, respectively. Among 130 low-mass
members in Tr 37, 63 stars are Class II or stars with transition disks.
The median age of Tr 37 in Table 2 of \citet{saa06} is 2.6 Myr.
The fraction of primordial disk is about 49 ($\pm$ 7) \%.
This value is slightly higher than other clusters of similar age.  However,
there are some active regions of current star-formation in this region 
(IC 1396A and IC 1396N), near the O6V star HD 206267.
This indicates that there are stars with a spread of age in this region,
which could bias the disk fraction to higher values - or, equivalently,
where the age based on optical photometry could be older than the average
age of all the young stars in the region (because the youngest stars are
embedded).
In NGC 7160 only 1 star among 14 low-mass members is classified as a CTTS.

\subsubsection{NGC 2362 \& Upp Sco}

The frequency of disked stars in the young open cluster NGC 2362 is well studied
by \citet{dh07}. They selected members from X-ray observation and spectroscopy.
There are 168 low-mass members among 232 members, and only 13 out of 168 stars
have primordial disks. The age of NGC 2362 is assumed to be 5 Myr, but a slightly
younger age is more favored from their Figure 2. We adopt the age of the cluster
as 4 ($\pm$ 1) Myr. \citet{cmhm06} published {\it Spitzer} IRAC 4.5 $\mu m$,
8.0 $\mu m$, and IRS 16 $\mu m$ data for 204 stars in Upper Scorpius (Upp Sco) OB 
association. Among them 85 stars belong to low-mass regime, and 12
out of 85 low-mass stars show an IR excess. The fraction of IR excess stars is
about 14 ($\pm$ 4) \%. This value is slightly higher than that of NGC 2362.
This could result from selection effects because Upp Sco is
a part of Sco-Cen OB association and therefore
the membership selection of low-mass stars could be biased toward X-ray
bright stars for the surveyed regions.

\subsubsection{$\eta$ Cha}

$\eta$ Cha is an interesting young stellar group of age 6 (-1, +2) Myr
\citep{ls04}. There are 18 known members in the group, and 7 are
low-mass stars (Sp = K7 -- M4). Among 7 low-mass stars, 2 stars show
an excess emission from {\it Spitzer} observation. Although the fraction
of stars with primordial disk is very high (about 29 $\pm$ 23 \%),
the number of member stars is very small. In addition, the spectral type of
the earliest star $\eta$ Cha is B8V, and the UV radiation field is therefore
far weaker than that of the other SFRs discussed here. The group is far
away from galactic disk. Such differences in environment could affect
the lifetime of disks of stars in $\eta$ Cha. 

\subsubsection{disk fraction versus age}

\begin{figure}
\epsscale{0.5}
\plotone{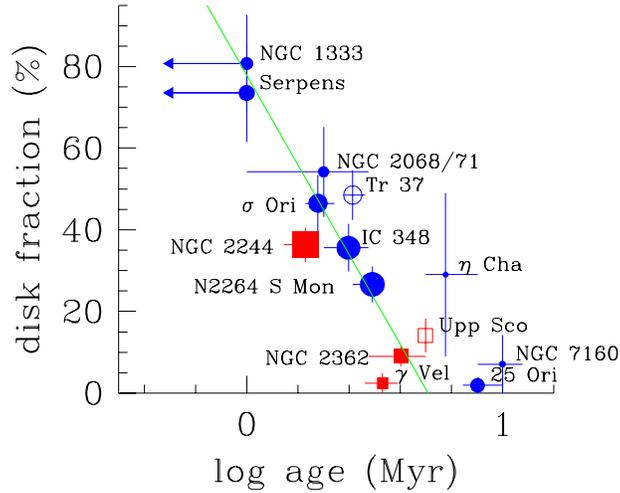}
\caption{The disk fraction versus age diagram. The primordial disk fraction
of young open clusters and groups is calculated for low-mass stars (m $\leq
1 M_\odot$) down to the completeness limit of membership selection.
A square represents the clusters containing stars with earliest spectral type of O7
or earlier or with evolved similar massive stars.
The solid line represents the relation between primordial disk fraction
and age of clusters younger than 6 Myr. We used a weighted regression (weight
= square root of the number of low-mass stars). Half of the normal weight was applied
for Tr 37 and Upp Sco (open symbols) due to the reasons mentioned in the text. Two young
clusters (NGC 2244, and $\gamma$ Vel) were omitted from the regression because
the disk fraction of these clusters is suspected to be strongly affected by
the presence of massive star(s). \label{diskevol} }
\end{figure}

The Taurus-Auriga SFR is another of the important SFRs that has been observed with {\it Spitzer} IRAC
\citep{lh05}. We did not try to re-analyze YSOs in Tau SFR because 
Tau SFR can be subdivided into several small SFRs that may have
different star formation histories.
The primordial disk fraction and age of each cluster are summarized in
Table \ref{tab_disk}, and is drawn in Figure \ref{diskevol}.

\placetable{tab_disk}
\placefigure{diskevol}

Figure \ref{diskevol} shows that the primordial disk fraction of NGC 2244 
and $\gamma$ Vel is slightly lower than the other clusters for a similar age.
And although
the most massive O7V star in NGC 2264 S Mon may affect the disks of low-mass
stars around the star (see \S 6.1), the overall disk fraction seems to be not
affected strongly by the presence of the O7V star (about 30 M$_\odot$).
In addition, for older clusters (age $\geq$ 6 Myr) the fraction of
primordial disks is very low, but not zero. Several H$\alpha$ emission
stars can be found in the 7.5 Myr cluster NGC 6531 (M21) \citep{psk01}.

The solid line in Figure \ref{diskevol} is the result from a weighted
regression. Two young clusters (NGC 2244 \& $\gamma$ Vel) and three
old (age $\geq$ 6 Myr) group and clusters were excluded from the regression.
The weights applied were proportional to the square-root of the number of low-mass members, and
that of Tr 37 and Upp Sco was half of the normal weight because of the reason
mentioned above. The regression result is

$$ f_{pdisk} = 78.0 (\pm 2.2) - 110. (\pm 5.0) \log {\rm age (Myr)} $$.

This result implies that the primordial disk fraction at birth is
about 80\% and the disk lifetime is 5.1 ($\pm$ 0.5) Myr.
The fraction of primordial disks in NGC 2244 and $\gamma$ Vel
is about 17\% lower than other clusters. And that of $\eta$ Cha is
higher than the others.
A common disk lifetime of about 5 Myr also indicates that planets 
should form (or at least mostly form) within this period. 
The disk lifetimes obtained here are consistent
with the values obtained by \citet{ssecs} from the observation of T Tauri stars
in the Tau-Aur SFR.

\section{DISCUSSION}

\subsection{Stars below the PMS Locus}

\citet{ghh54} pointed out that the Ae star (or now possibly a Herbig Be star -
\citet{pmvj08}) W90 is 3 mag fainter than normal A type stars. Later
\citet{ssycg72} proposed that the location in the CMD could be explained
by assuming that it was being viewed through a nearly edge-on circumstellar
disk/envelope containing primarily large dust grains. On the other hand,
\citet{ps02} suggested a prolonged period of star formation from 
the presence of some fainter PMS stars in the CMDs of the Trapezium cluster and
the Taurus SFR. \citet{shc04} found a population
of faint PMS stars in the Trapezium cluster whose inferred age would be of
order 10 Myr if judged directly from isochrone fitting, they concluded
that these stars were best explained by the disk obscuration model.
From the spatial density distribution and their near-IR excess, \citet{sbcki08}
supported the near edge-on disk model of below the PMS locus (BMS) stars.

\begin{figure}
\epsscale{0.5}
\plotone{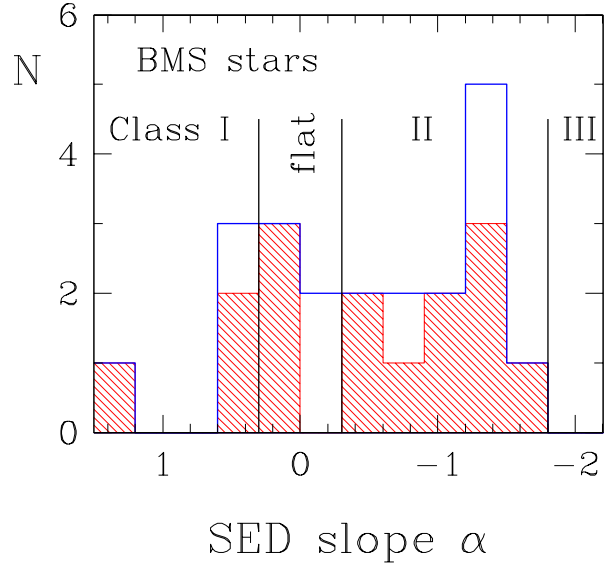}
\caption{The distribution of the SED slope $\alpha$ of BMS stars.
\label{figbms} }
\end{figure}

\begin{figure}
\epsscale{0.5}
\plotone{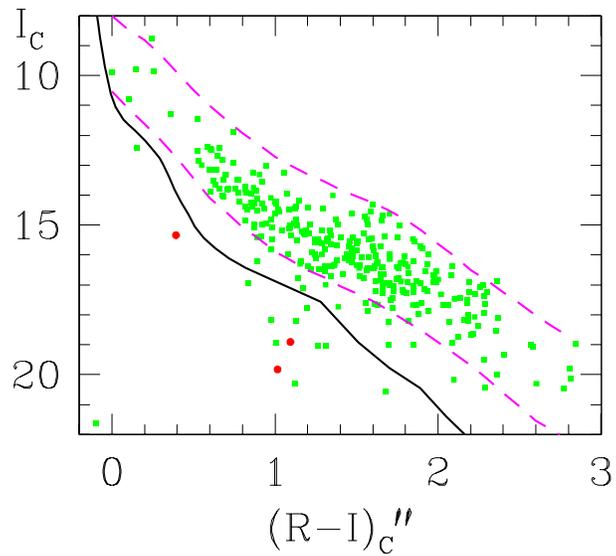}
\caption{The $I_C$ versus ($R-I$)$_C''$ color-magnitude diagram of
Class II objects and three Class I objects classified as BMS candidates.
The dashed lines represent the locus of PMS stars in NGC 2264,
while the solid line denotes the reddened zero-age main sequence relation
shifted to the distance of NGC 2264. \label{figbmscmd} }
\end{figure}

\placefigure{figbmscmd}
\placefigure{figbms}

\citet{sbcki08} presented 82 BMS candidates. Among them, 15 were outside
the {\it Spitzer} field-of-view, 8 were double in IRAC, 11 were not detected,
2 were detected only in [3.6] and 22 were detected only in [3.6] and [4.5].
There were only 24 objects that were detected in 3 or more channels and whose
SED slope $\alpha$ could be calculated reliably. 
(We force fitted a line to the SED although the data did not meet the 
criteria mentioned in \S 3.2). The SED slope
distribution of BMS stars is presented in Figure \ref{figbms}.
Most of the BMS stars (21/24) were disked stars (Class I, flat, or Class II).
Three stars (C39036 = LkH$\alpha$ 71, C35777,
and C48145) show the SED of stellar photospheres ($\alpha$ = -2.6 -- -2.8).
Ogura 97 (= C32005 = SST 10710), although the star is not
classified as a BMS star in \citet{sbcki08}, is a BMS star as H$\alpha$
emission has been detected by \citet{ko84}; it is a Class I object.
With 21 disks and a total of 59 objects where we could have reliably detected
a disk (82 - 15 - 8), the fraction of BMS objects with primordial 
disks is about 36\%. This value is
slightly higher than the primordial disk fraction of members in the PMS locus.
As we did not impose any limitation in brightness, the fraction obtained above
is a minimum value.
If we restrict BMS membership to those candidates with strong 
H$\alpha$ emission (membership class:
``+'' or ``H''), the fraction increases to about 55\% (= 18/33).\footnote{As
the bright, spatially varying nebulosity in NGC 2264 could affect the quality
of photometry, we cannot rule out the possibility of spurious detections of 
H$\alpha$ emission of some BMS candidates, particularly the fainter ones. Actually the BMS
candidates not detected in IRAC or only detected in [3.6] and/or [4.5] are
mostly faint ($I_C >$ 18 mag for strong H$\alpha$ emission stars and
$I_C >$ 17.5 mag for weak H$\alpha$ emission stars).}

There are some stars that could be classified as BMS stars (Class I or II)
in Figure \ref{figopt}. We list 10 new BMS candidates in Table \ref{tab_bms}
and drew the color-magnitude diagram of Class II
objects and three BMS candidate Class I objects in Figure \ref{figbmscmd}. 
Three were not listed in Table 12 of \citet{sbcki08} because of
small values of $\Delta I_C$ ($\Delta I_C <$ 0.5 mag), three were too
faint ($I_C >$ 20 mag) and four had no signature of H$\alpha$ emission. 
But most of the faint Class I objects are not listed in Table \ref{tab_bms}
because of their optical images and their spatial location (Field or Halo)
- two (SST 3145 \& 7707) are too faint to discern whether they are faint stars
or galaxies and two (SST 9861 \& 21808) are in the vicinity of faint galaxies
(see \S2.5).

\placetable{tab_bms}

The high disk fraction of BMS stars strongly suggests that
BMS stars are bona fide YSOs. In addition, a well-known young disked K7 WTTS
star KH 15D in NGC 2264 is close to the lower boundary of the PMS locus of
NGC 2264 during its bright phase, but becomes a BMS star
during eclipse. The star is fainter by about 3.5 mag in $I$ during eclipse,
but is bluer by about 0.2 mag in $V-I$ \citep{cmh05} probably due to
the contribution of scattered light. Although the H$\alpha$ emission of
KH 15D is not detected photometrically, \citet{ds05} reported
an H$\alpha$ emission equivalent width of 103.1$\AA$. 
In addition, the spectra of KH 15D during eclipse do not show any change
in spectral type or reddening, indicating that the extinction is caused by
larger dust grains \citep{hhsf01,abwc04}. And the polarization dramatically
increased from practically $\sim$ 0\% during out of eclipse to $\sim$ 2\%
during eclipse \citep{abwc04}. In addition the variation amplitude in [3.6]
of KH 15D is nearly the same as that in $I_C$ (see \S 7.2.5), also indicative
of gray extinction by large grains.  KH15D therefore provides direct
evidence in favor of the disk extinction model to explain BMS stars.

\subsection{Interesting Objects}

\subsubsection{W90 (= S2144)}

The well known BMS star W90 (= S2144 = SST 9597) was first found by
\citet{ghh54} as an emission line A type star. He also pointed out
that it is fainter by 3 magnitudes relative to other PMS stars of 
similar spectral type in the cluster.
Its PMS nature was confirmed spectroscopically by \citet{ssy71}.
From photometric studies, \citet{sbl97,sbcki08} found many
BMS stars in the cluster. Recently \citet{pmvj08} published
a vast amount of archival and original photometric and spectroscopic
data for W90. They showed the variability and spectral changes of the star.

As mentioned in the previous section, W90 is very red in [5.8]-[8.0].
W90 could be classified as either a Class I or Class II object. We classified
the star as Class II due to its relatively bluer color in [8.0]-[24].
We present the spectral energy distribution of W90 in Figure \ref{figsed}.
W90 shows excess emission in the near-IR as well as in all IRAC channels.

\subsubsection{C35527 - A Class I Object in the PMS locus}

\begin{figure}
\epsscale{0.5}
\plotone{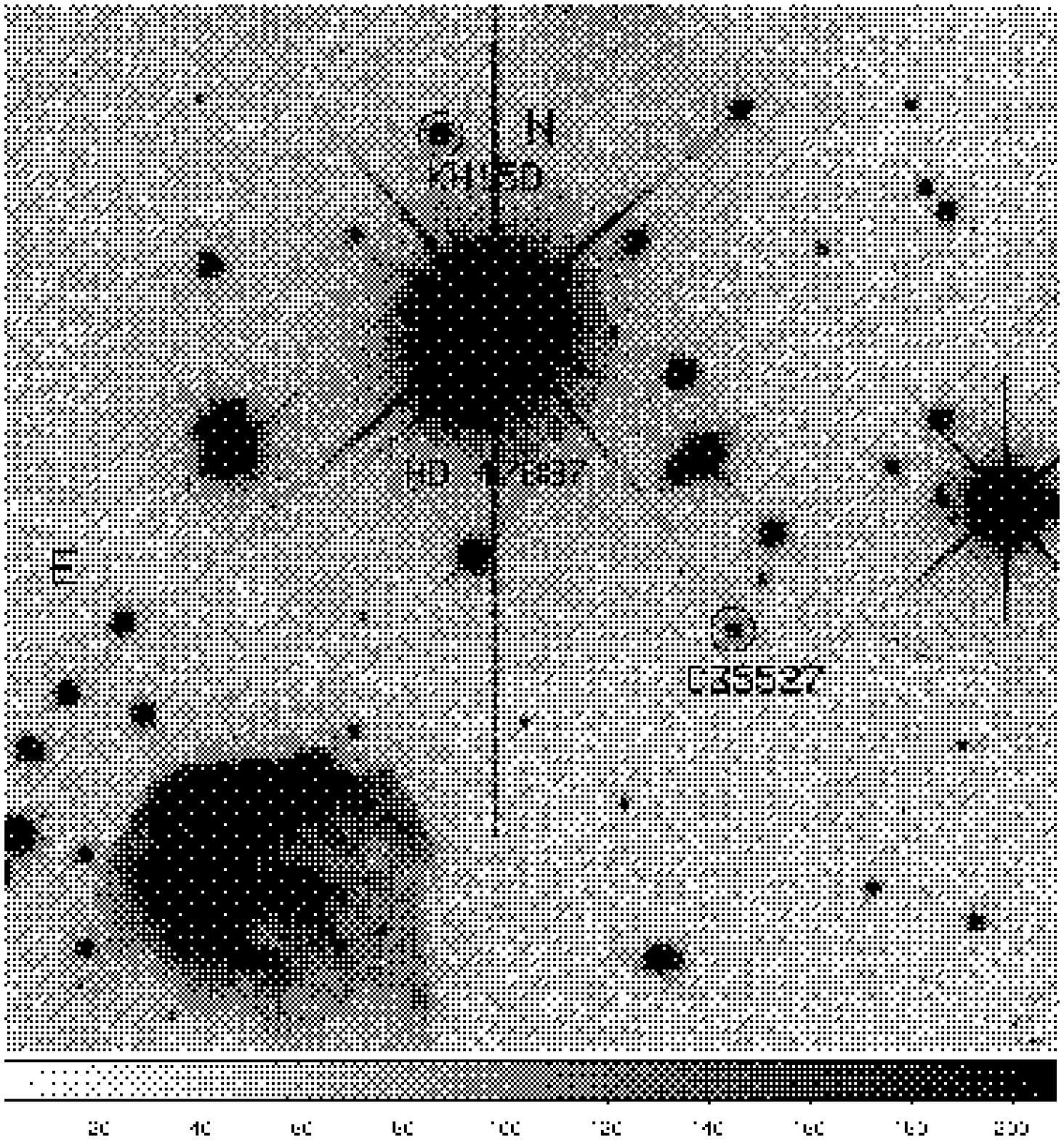}
\caption{The finder chart of the Class I star C35527. Thin nebulosity can
be seen around the star.
\label{figc35527} }
\end{figure}

C35527 (= SST 12918; $\alpha_{\rm J2000}$ = 6$^h$ 41$^m$ 06.$^s$44,
$\delta_{\rm J2000}$ = 9$^\circ$ 36$'$ 58.$''$6) is the only Class I object
in the PMS locus. The star is 1.$'$85 SW of KH 15D (see Figure \ref{figc35527}),
and is barely detected
in the digital sky survey. But in our CFHT images we can see the object clearly.
As can be seen in Figure \ref{figc35527}, the star is surrounded by a thin
nebulosity. The star is a strong X-ray source (detection significance = 75.05).
Its 2MASS colors are those of a highly reddened normal star. 
Even in $K_S$-[3.6],
the star does not show any signature of an IR excess (Figure \ref{fig2ms}
(upper left) or Figure \ref{figsed} (upper right)). This may be caused in part by
the variability of the object. But in $K_S$-[4.5] or longer wavelengths,
the star shows an evident IR excess, which increases to 
longer wavelength. 

\placefigure{figc35527}

The star was detected in our SSO $I_C$ band image (Date of observation: 1997
November 19, UT = 16:15:47.5, MJD = 50771.677627; no ``S'' ID was assigned).
Its $I_C$ magnitude was 17.765 $\pm$ 0.068 (N$_{\rm obs}$ = 1). The magnitude and
colors from the observation with the CFHT are $I_C$ = 18.480, ($R-I$)$_C''$
= 2.368 and ($V-I$)$_C$ = 5.299. The star is very red and is variable. 

The spectral energy distribution of this star is shown in Figure \ref{figsed}.
As there is no direct method to determine the reddening of such a very red star,
we tried to fit the optical $VR_CI_C$ and 2MASS data to a model
atmosphere of effective temperature 2600K and surface gravity $\log g = 2.5$. 
The resulting reddening $E(B-V)$ is about 1.40 mag. The star shows
excess emission only in IRAC channels and in 24 $\mu$m.

\subsubsection{FU Ori type Candidate AR 6A (= C33859) and AR 6B}

\citet{ar03} suggested AR 6A (= C33859) is an FU Ori
type PMS star. IRAC colors, including MIPS [24], indicate the star is a Class II
object. From near-IR photometry and spectroscopy, \citet{ar03} estimated
the reddening and temperature ($A_V$ = 18 mag, $T_{BB}$ = 3000 K).
Although the star has very red colors ($(R-I)''$ = 3.09 $\pm$ 0.02),
the estimated reddening from its SED (see Figure \ref{figsed})
is $E(B-V) \approx$ 3.65 mag if we adopt
the optical spectral type of the star as G III (T$_{eff} \sim$ 5000 K)
suggested from spectral features in the K-band. This value is
much smaller than the value estimated by \citet{ar03}.
Such a large discrepancy may be caused by their using limited, near-IR
data which are affected by reddening as well as thermal emission, as they have
pointed out. In addition, \citet{sbcki08} noted that its location in the
near-IR C-C diagram cannot be simply interpreted as the result of reddening.

AR 6B (= SST 11837) was not detected in the deep optical survey of 
\citet{sbcki08},
but the IRAC colors, as well as the near-IR photometry of \citet{ar03}
suggest that it is a Class I object. The reddening
of AR 6B cannot be determined using the near-IR data mentioned above.
If we assume the same reddening as that of AR 6A, we cannot fit the SED
at all. If we assume $E(B-V)$ = 0.0, the blackbody flux of the object is
slightly fainter than the detection limit at $I_C$ ($\log \lambda F_\lambda
\approx$ -13.85 at $I_C$ = 22 mag).

\subsubsection{Allen's source}

\citet{daa72} discovered a strong IR source just north of the Cone nebula
and interpreted it as the most massive and luminous object in
NGC 2264. Currently the object is considered to be a deeply embedded, early-type
star lying within a massive molecular core \citep{sed08}.
This object is also known as NGC 2264 IRS1 (= C36261 = 2MASS J06411015+0929336,
$I_C$ = 19.068, $J$ = 11.508, $H$ = 7.643, $K_S$ = 4.924). Allen's source
is the brightest mid-IR source in NGC 2264 and is the only object saturated
in all short exposure IRAC images. \citet{mly89} obtained the SED slope
of the object ($\alpha$ = +0.1). Although there are several YSOs around IRS1,
the slope measured by \citet{mly89} is unlikely to be affected as Allen's
source is one of the brightest objects in the sky in mid-IR and is far
brighter than any of the neighboring objects.
We consider Allen's source to be a deeply embedded flat spectrum YSO.
A very strong, hard X-ray source (EXS-1 = SST 13380, T$_X \approx$ 100 MK)
lies 11$''$ southwest of Allen's source \citep{sd05}. The object is detected
in the short 3 channels of IRAC and therefore 
was not classified by its evolutionary stage. 
If we classify the object in the ([3.6]-[4.5], [4.5]-[5.8]) diagram, EXS-1
is a Class I object.

\subsubsection{The Variability of KH 15D}

One of the more interesting objects in NGC 2264, KH 15D (=C36305 = SST 13485).
KH 15D is very close to IRS1, the brightest IR source
in NGC 2264, and the star is on the bright spike of IRS1. 
It is detected only in channels 1 and 2 of AORs 3956736 and 3957248 
(phase = 0.31 of the eclipse light curve, \citet{cmh05}), 
AORs 3956480 and 3956992 (phase = 0.85 of the eclipse light
curve) is on the bright spike and
too faint to be measured.
In channels 3 and 4, the signal is too weak relative to the brightness of
the spike of NGC 2264 IRS1, and therefore could not be measured.
Its [3.6]-[4.5] color is slightly red, but not so red as to be classified
as a Class II object. Recently,
Marengo et al. (2009, in preparation) have obtained the light curve of KH 15D in
the four
IRAC bands, and confirmed the amplitude in mid-IR wavelengths is the same as
that in the optical, which implies that the opacity in the mid-IR is gray, and
therefore the size of dust grains in the circumbinary disk around
KH 15D is very large (see \citet{wh08}).

\section{SUMMARY}

We have performed mid-IR photometry of the young open cluster NGC 2264
surveyed with the IRAC and MIPS instruments of the {\it Spitzer} Space Telescope.
The results we obtained are summarized as follows:

(1) We presented the {\it Spitzer} IRAC [3.6], [4.5], [5.8], [8.0] and MIPS
[24] data of about 22,000 objects in NGC 2264. We also searched for
the optical, near-IR and X-ray counterparts of all mid-IR sources.

(2) We classified the YSOs in NGC 2264 using two classification 
schemes - one based on color-color diagrams, the other on the spectral energy
distributions. We introduced a parameter $Q_{CC}$ in the
photometric classification to make full use of the imformation from
multicolor photometry. We also compared the two classifications and found
them to be generally consistent with each other.

(3) From the spatial distributions of disked stars (Class I and II objects),
we identified two subclusterings of Class I objects in the CONE region
of \citet{sbcki08}. One was the well-known Spokes cluster \citep{pst06};
the other was located near the head of the Cone nebula.
On the other hand, the S MON region of \citet{sbcki08} mostly comprised 
Class II objects. These subclusterings in NGC 2264 show a distinct difference
in the distribution of SED slopes and the mean value of the SED slopes.

(4) We examined the disk fraction of below the PMS locus (BMS) stars
and found that the fraction with
primordial disks to be about 35\%. This value is higher
than that of normal stars in the PMS locus and supports 
the idea that BMS stars are young stars with nearly edge-on disks.

(5) We also obtained the result that the fraction of primordial disks increases
as the distance from the most massive star S Mon (O7V) increases.
This is an important clue that the UV radiation and winds of massive stars
affects the fate of primordial disks of nearby low-mass stars.

(6) From a re-analysis of data for young open clusters observed
with the {\it Spitzer} space telescope, we found that a common disk lifetime
was about 5 Myr, the disk fraction of
low-mass stars in NGC 2244 and $\gamma$ Vel was slightly lower than other
clusters with a similar age, and that of $\eta$ Cha, although the error is large,
seems to be higher. This fact implies that the strength of the UV radiation field
is one of the most important factors affecting the fate of disks around low-mass stars.

\acknowledgments 
H.S. would like to express his deep thanks to the {\it Spitzer} Science Center
for hosting him as a Visiting Associate for a year.  He also thanks
J.-E. Lee for valuable suggestions.
H.S. acknowledges the support of the Korea Science and Engineering
Foundation (KOSEF) to the Astrophysical Research Center for
the Structure and Evolution of the Cosmos (ARCSEC$''$) at Sejong University.

\clearpage

\clearpage

\begin{deluxetable}{rcccccccccccccccccccccl}
\tablecolumns{23}
\tabletypesize{\scriptsize}
\rotate
\tablecaption{Catalogue of Sources Detected with {\it Spitzer} Space Telescope IRAC and MIPS 24$\mu$m \tablenotemark{b} \label{tab_sst}}
\tablewidth{0pt}
\tablehead{
\colhead{SST} & \colhead{$\alpha_{J2000}$} & \colhead{$\delta_{J2000}$} & \colhead{[3.6]} &
\colhead{[4.5]} & \colhead{[5.8]} & \colhead{[8.0]} & \colhead{[24]} &
\colhead{$\epsilon_{[3.6]}$} & \colhead{$\epsilon_{[4.5]}$} & \colhead{$\epsilon_{[5.8]}$} &
\colhead{$\epsilon_{[8.0]}$} & \colhead{$\epsilon_{[24]}$} & \multicolumn{5}{c}{N$_{obs}$} &
\colhead{dup\tablenotemark{b}} & \colhead{memb\tablenotemark{c}} & \colhead{2MASS} & \colhead{Sung et al. (2008)} }

\startdata
 10936&  6:40:53.63 &   9:40:09.2 &  15.938 & 15.907 &\nodata &\nodata &\nodata &  0.064 &  0.049 &\nodata &\nodata &\nodata & 4 & 4 & 0 & 0 & 0 &  &       &                  &                      \\
 10937&  6:40:53.63 &   9:29:53.3 &  13.171 & 12.908 & 12.561 & 11.891 &\nodata &  0.043 &  0.058 &  0.043 &  0.016 &\nodata & 7 & 8 & 4 & 4 & 0 &  &  II,H & 06405363+0929532 & C32434               \\
 10938&  6:40:53.63 &   9:47:04.5 &  11.887 & 11.621 & 11.333 & 10.857 &  8.404 &  0.015 &  0.040 &  0.054 &  0.042 &  0.300 & 8 & 8 & 7 & 8 & 1 &D &  II,M & 06405362+0947043 & C32440+C32444        \\
 10939&  6:40:53.64 &   9:33:24.7 &  10.757 & 10.558 & 10.404 & 10.194 &  6.880 &  0.018 &  0.020 &  0.021 &  0.036 &  0.064 & 8 & 8 & 8 & 8 & 1 &  & II/III,+ & 06405363+0933247 & S2564                \\
 10940&  6:40:53.64 &   9:36:46.1 &  14.787 & 14.734 & 14.409 &\nodata &\nodata &  0.016 &  0.031 &  0.069 &\nodata &\nodata & 4 & 4 & 2 & 0 & 0 &  & X     & 06405362+0936461 & C32439               \\
 10941&  6:40:53.64 &   9:34:36.7 &  15.726 & 15.311 &\nodata &\nodata &\nodata &  0.038 &  0.059 &\nodata &\nodata &\nodata & 4 & 4 & 0 & 0 & 0 &  &       &                  &                      \\
 10942&  6:40:53.64 &  10:01:08.9 &  15.094 & 15.008 &\nodata &\nodata &\nodata &  0.059 &  0.096 &\nodata &\nodata &\nodata & 2 & 2 & 0 & 0 & 0 &  &       & 06405364+1001087 & C32446               \\
 10943&  6:40:53.66 &   9:22:10.0 &  16.488 & 16.300 &\nodata &\nodata &\nodata &  0.086 &  0.080 &\nodata &\nodata &\nodata & 4 & 3 & 0 & 0 & 0 &  &       &                  &                      \\
 10944&  6:40:53.68 &   9:23:34.7 &  15.845 & 15.792 &\nodata &\nodata &\nodata &  0.062 &  0.032 &\nodata &\nodata &\nodata & 4 & 4 & 0 & 0 & 0 &D &       &                  & C32445+C32458        \\
 10945&  6:40:53.67 &   9:58:00.1 &  12.730 & 12.645 & 12.665 & 12.655 &\nodata &  0.032 &  0.044 &  0.032 &  0.095 &\nodata & 8 & 8 & 4 & 4 & 0 &  & M     & 06405367+0958000 & C32453               \\
 10946&  6:40:53.68 &   9:31:35.3 &  16.741 &\nodata &\nodata &\nodata &\nodata &  0.006 &\nodata &\nodata &\nodata &\nodata & 2 & 0 & 0 & 0 & 0 &  &       &                  &                      \\
 10947&  6:40:53.71 &   9:19:38.3 &  16.014 & 16.192 &\nodata &\nodata &\nodata &  0.027 &  0.071 &\nodata &\nodata &\nodata & 4 & 4 & 0 & 0 & 0 &  &       &                  & C32463               \\
 10948&  6:40:53.72 &   9:18:53.2 &  17.102 &\nodata &\nodata &\nodata &\nodata &  0.087 &\nodata &\nodata &\nodata &\nodata & 2 & 0 & 0 & 0 & 0 &  &       &                  & C32462               \\
 10949&  6:40:53.72 &   9:25:56.5 &  16.383 &\nodata &\nodata &\nodata &\nodata &  0.047 &\nodata &\nodata &\nodata &\nodata & 4 & 0 & 0 & 0 & 0 &  &       &                  & C32461               \\
 10950&  6:40:53.73 &   9:39:16.1 &  16.435 & 16.100 &\nodata &\nodata &\nodata &  0.086 &  0.040 &\nodata &\nodata &\nodata & 4 & 3 & 0 & 0 & 0 &  &       &                  &                      \\
 10951&  6:40:53.73 &   9:46:20.8 &  14.853 & 14.717 &\nodata &\nodata &\nodata &  0.025 &  0.075 &\nodata &\nodata &\nodata & 4 & 4 & 0 & 0 & 0 &  &       &                  & C32468               \\
 10952&  6:40:53.74 &   9:46:37.7 & \nodata & 15.976 &\nodata &\nodata &\nodata &\nodata &  0.106 &\nodata &\nodata &\nodata & 0 & 3 & 0 & 0 & 0 &  &       &                  &                      \\
 10953&  6:40:53.76 &   9:53:46.0 &  16.053 & 16.114 &\nodata &\nodata &\nodata &  0.084 &  0.016 &\nodata &\nodata &\nodata & 4 & 2 & 0 & 0 & 0 &  &       &                  & C32470               \\
 10954&  6:40:53.78 &   9:30:39.0 &  10.232 & 10.246 & 10.202 & 10.213 &\nodata &  0.015 &  0.008 &  0.009 &  0.020 &\nodata & 8 & 8 & 8 & 8 & 0 &  & X     & 06405377+0930389 & S2568                \\
\enddata

\tablenotetext{a}{Table \ref{tab_sst} is presented in its entirety in the electronic
edition of the Astronomical Journal. A portion is shown here for guidance
regarding its form and content. Units of right ascension are hours, minutes,
and seconds of time, and units of declination are degrees, arcminutes,
and arcseconds.}
\tablenotetext{b}{duplicity - D: SST source having two optical counterparts within 2$''$ searching radii, T: SST source having three optical counterparts within 2$''$ searching radii.}
\tablenotetext{c}{membership - IR classification (Class I, II, II/III, pre-TD (pre-transition disk), and TD(transition disk)) and other membership listed in Sung et al. (2008)}
\end{deluxetable}

\begin{deluxetable}{rccccccccccccccccccccl}
\tablecolumns{22}
\tabletypesize{\scriptsize}
\rotate
\tablecaption{{\it Spitzer} Counterparts of X-Ray Sources without Optical or 2MASS counterparts \label{tab_Xray}}
\tablewidth{0pt}
\tablehead{
\colhead{SST} & \colhead{$\alpha_{J2000}$} & \colhead{$\delta_{J2000}$} & \colhead{[3.6]} &
\colhead{[4.5]} & \colhead{[5.8]} & \colhead{[8.0]} & \colhead{[24]} &
\colhead{$\epsilon_{[3.6]}$} & \colhead{$\epsilon_{[4.5]}$} & \colhead{$\epsilon_{[5.8]}$} &
\colhead{$\epsilon_{[8.0]}$} & \colhead{$\epsilon_{[24]}$} & \multicolumn{5}{c}{N$_{obs}$} &
\colhead{Class} & \colhead{Signif\tablenotemark{a}} & \colhead{dist($''$)} }

\startdata
  5721&  6:40:20.12 &   9:45:10.9 &  14.762 & 14.054 & 13.672 & 11.892 &  8.523 &  0.029 &  0.029 &  0.080 &  0.060 &  0.262 & 4 & 4 & 4 & 4 & 1 & Galaxy& 25.4& 0.3 \\
  6944&  6:40:27.73 &   9:41:19.8 &  16.595 & 15.875 &\nodata &\nodata &\nodata &  0.075 &  0.025 &\nodata &\nodata &\nodata & 2 & 4 & 0 & 0 & 0 &       & 7.8 & 0.9 \\
  7546&  6:40:31.61 &   9:44:26.8 &  15.537 & 15.115 &\nodata &\nodata &\nodata &  0.037 &  0.047 &\nodata &\nodata &\nodata & 4 & 4 & 0 & 0 & 0 &       & 8.8 & 0.4 \\
  8077&  6:40:34.90 &   9:44:23.2 &  16.174 & 15.727 &\nodata &\nodata &\nodata &  0.525 &  0.036 &\nodata &\nodata &\nodata & 1 & 4 & 0 & 0 & 0 &       & 5.2 & 0.8 \\
  8097&  6:40:35.04 &   9:39:54.4 &  15.567 & 15.412 &\nodata &\nodata &\nodata &  0.040 &  0.056 &\nodata &\nodata &\nodata & 4 & 4 & 0 & 0 & 0 &       & 7.0 & 0.7 \\
  8221&  6:40:35.81 &   9:52:00.5 &  14.768 & 14.733 &\nodata &\nodata &\nodata &  0.038 &  0.020 &\nodata &\nodata &\nodata & 2 & 2 & 0 & 0 & 0 &       & 6.3 & 1.0 \\
  8614&  6:40:38.16 &   9:32:01.4 &  16.110 & 14.734 &\nodata &\nodata &\nodata &  0.056 &  0.642 &\nodata &\nodata &\nodata & 4 & 4 & 0 & 0 & 0 &       & 4.9 & 0.3   \\
  8715&  6:40:38.77 &   9:51:44.8 &  15.351 & 15.205 &\nodata &\nodata &\nodata &  0.051 &  0.019 &\nodata &\nodata &\nodata & 4 & 4 & 0 & 0 & 0 &       & 4.2 & 1.1   \\
  9444&  6:40:43.62 &   9:31:14.9 &  16.028 & 15.623 &\nodata &\nodata &\nodata &  0.079 &  0.053 &\nodata &\nodata &\nodata & 4 & 4 & 0 & 0 & 0 &       & 4.3 & 1.3   \\
  9735&  6:40:45.49 &   9:37:06.7 &  16.572 & 16.167 &\nodata &\nodata &\nodata &  0.009 &  0.050 &\nodata &\nodata &\nodata & 4 & 2 & 0 & 0 & 0 &       & 4.2 & 0.8   \\
  9934&  6:40:46.83 &   9:32:16.8 & \nodata & 15.763 &\nodata &\nodata &\nodata &\nodata &  0.110 &\nodata &\nodata &\nodata & 0 & 4 & 0 & 0 & 0 &       & 4.4 & 0.4   \\
 10274&  6:40:49.09 &   9:31:56.5 &  16.290 & 16.238 &\nodata &\nodata &\nodata &  0.023 &  0.441 &\nodata &\nodata &\nodata & 3 & 1 & 0 & 0 & 0 &       & 3.8 & 0.5   \\
 10453&  6:40:50.28 &   9:33:11.3 &  15.310 & 15.049 &\nodata &\nodata &\nodata &  0.023 &  0.051 &\nodata &\nodata &\nodata & 4 & 4 & 0 & 0 & 0 &       & 4.1 & 0.8   \\
 10493&  6:40:50.47 &   9:32:19.6 & \nodata & 16.150 &\nodata &\nodata &\nodata &\nodata &  0.087 &\nodata &\nodata &\nodata & 0 & 3 & 0 & 0 & 0 &       & 20.8& 0.5   \\
 10858&  6:40:52.97 &   9:32:43.2 &  16.134 & 15.710 &\nodata &\nodata &\nodata &  0.061 &  0.061 &\nodata &\nodata &\nodata & 4 & 4 & 0 & 0 & 0 &       & 4.3 & 0.8   \\
 11102&  6:40:54.67 &   9:53:04.4 &  16.310 & 15.529 &\nodata &\nodata &\nodata &  0.018 &  0.055 &\nodata &\nodata &\nodata & 2 & 4 & 0 & 0 & 0 &       & 4.7 & 0.6   \\
 11257&  6:40:55.77 &   9:54:11.5 & \nodata & 15.928 &\nodata &\nodata &\nodata &\nodata &  0.015 &\nodata &\nodata &\nodata & 0 & 2 & 0 & 0 & 0 &       & 3.5 & 0.6   \\
 11315&  6:40:56.17 &   9:37:55.1 &  15.936 & 15.584 &\nodata &\nodata &\nodata &  0.065 &  0.034 &\nodata &\nodata &\nodata & 4 & 4 & 0 & 0 & 0 &       & 4.0 & 1.3   \\
 11592&  6:40:57.98 &   9:36:39.5 &  11.279 &  9.541 &  8.540 &  7.645 &  4.040 &  0.064 &  0.062 &  0.051 &  0.039 &  0.013 & 8 & 8 & 8 & 8 & 1 &   I   & 5.4 & 0.4   \\
 11598&  6:40:58.01 &   9:36:14.6 &  12.892 & 11.805 & 11.077 & 10.341 &  6.682 &  0.028 &  0.038 &  0.031 &  0.014 &  0.066 & 8 & 8 & 8 & 8 & 1 &   I   & 4.1 & 0.3   \\
 11711&  6:40:58.68 &   9:52:53.6 &  13.978 & 13.928 & 13.695 & 12.441 &\nodata &  0.031 &  0.044 &  0.067 &  0.334 &\nodata & 5 & 4 & 3 & 1 & 0 &       & 7.3 & 0.0   \\
 11713&  6:40:58.70 &   9:54:16.1 &  13.286 & 13.160 & 12.925 &\nodata &\nodata &  0.041 &  0.105 &  0.058 &\nodata &\nodata & 8 & 6 & 4 & 0 & 0 &       & 9.6 & 0.4   \\
 12412&  6:41:03.21 &   9:44:44.2 &  16.208 & 15.272 &\nodata &\nodata &\nodata &  0.044 &  0.068 &\nodata &\nodata &\nodata & 4 & 4 & 0 & 0 & 0 &       & 5.3 & 1.0   \\
 12637&  6:41:04.61 &   9:36:18.3 &  10.832 &  9.374 &  8.619 &  7.950 &\nodata &  0.035 &  0.028 &  0.012 &  0.023 &\nodata & 8 & 8 & 8 & 8 & 0 &   I   & 40.1& 0.4   \\
 12681&  6:41:05.01 &   9:41:04.7 &  16.289 & 16.078 &\nodata &\nodata &\nodata &  0.043 &  0.074 &\nodata &\nodata &\nodata & 4 & 4 & 0 & 0 & 0 &       & 4.0 & 0.5   \\
 12722&  6:41:05.23 &   9:36:31.9 &  12.252 & 11.522 & 10.110 &\nodata &\nodata &  0.098 &  0.104 &  0.119 &\nodata &\nodata & 1 & 8 & 8 & 0 & 0 &       & 5.8 & 0.5   \\
 12774&  6:41:05.53 &   9:35:01.4 &  14.083 & 13.732 & 13.221 & 12.694 &\nodata &  0.033 &  0.059 &  0.069 &  0.014 &\nodata & 4 & 4 & 4 & 2 & 0 &  II   & 4.1 & 0.2   \\
 12780&  6:41:05.56 &   9:34:08.0 &  12.312 & 10.675 &  9.612 &  8.809 &  3.544 &  0.033 &  0.047 &  0.038 &  0.012 &  0.100 & 4 & 3 & 8 & 8 & 1 &   I   & 5.9 & 0.9   \\
 12831&  6:41:05.87 &   9:34:46.0 &  11.427 & 11.281 & 10.855 & 10.996 &\nodata &  0.026 &  0.014 &  0.078 &  0.097 &\nodata & 8 & 8 & 8 & 8 & 0 & II/III& 65.4& 0.4   \\
 12875&  6:41:06.19 &   9:34:08.8 &  12.664 & 10.604 &  9.831 &  9.255 &  3.576 &  0.060 &  0.030 &  0.021 &  0.019 &  0.072 & 8 & 8 & 8 & 8 & 1 &   I   & 3.7 & 0.6   \\
 13029&  6:41:07.16 &   9:30:36.3 &  11.195 & 10.637 & 10.097 &  9.457 &  6.003 &  0.060 &  0.036 &  0.043 &  0.017 &  0.100 & 8 & 8 & 8 & 8 & 1 &  II   & 13.3& 1.0   \\
 13085&  6:41:07.47 &   9:36:10.8 &  14.458 & 13.612 & 13.114 &\nodata &\nodata &  0.049 &  0.028 &  0.086 &\nodata &\nodata & 4 & 4 & 4 & 0 & 0 &       & 3.9 & 1.1   \\
 13111&  6:41:07.67 &   9:34:19.1 &  13.621 & 11.252 &  9.980 &  9.185 &  4.095 &  0.070 &  0.022 &  0.020 &  0.013 &  0.019 & 8 & 8 & 8 & 8 & 1 &   I   & 34.4& 0.5   \\
 13672&  6:41:11.61 &   9:29:10.3 &  13.291 & 11.254 & 10.135 &\nodata &\nodata &  0.113 &  0.063 &  0.072 &\nodata &\nodata & 6 & 8 & 7 & 0 & 0 &       & 5.6 & 0.6   \\
 13749&  6:41:12.22 &   9:29:14.5 &  12.249 & 10.869 &\nodata &\nodata &\nodata &  0.188 &  0.170 &\nodata &\nodata &\nodata & 7 & 8 & 0 & 0 & 0 &       & 6.7 & 0.3   \\
 14094&  6:41:14.49 &   9:35:37.5 &  15.894 & 14.534 &\nodata &\nodata &\nodata &  0.047 &  0.055 &\nodata &\nodata &\nodata & 3 & 3 & 0 & 0 & 0 &       & 3.6 & 0.9   \\
 14144&  6:41:14.84 &   9:29:17.3 &  15.650 & 14.758 &\nodata &\nodata &\nodata &  0.070 &  0.355 &\nodata &\nodata &\nodata & 4 & 1 & 0 & 0 & 0 &       & 3.7 & 0.7   \\
 14338&  6:41:16.41 &   9:52:49.7 &  16.582 & 15.482 &\nodata &\nodata &\nodata &  0.095 &  0.044 &\nodata &\nodata &\nodata & 4 & 4 & 0 & 0 & 0 &       & 10.2& 0.8   \\
 14371&  6:41:16.65 &   9:52:06.8 & \nodata & 16.535 &\nodata &\nodata &\nodata &\nodata &  0.067 &\nodata &\nodata &\nodata & 0 & 2 & 0 & 0 & 0 &       & 6.9 & 0.4  \\
 14834&  6:41:19.72 &   9:31:39.0 &  15.198 & 15.063 &\nodata &\nodata &\nodata &  0.052 &  0.038 &\nodata &\nodata &\nodata & 4 & 4 & 0 & 0 & 0 &       & 6.2 & 1.0   \\
 15746&  6:41:25.79 &   9:40:35.5 &  16.651 & 16.065 &\nodata &\nodata &\nodata &  0.055 &  0.188 &\nodata &\nodata &\nodata & 4 & 3 & 0 & 0 & 0 &       & 7.0 & 1.0   \\
 16270&  6:41:29.08 &   9:39:49.4 &  16.156 & 15.479 &\nodata &\nodata &\nodata &  0.023 &  0.041 &\nodata &\nodata &\nodata & 4 & 4 & 0 & 0 & 0 &       & 11.9& 1.0   \\
 16562&  6:41:31.03 &   9:35:25.4 &  16.335 & 15.506 &\nodata &\nodata &\nodata &  0.036 &  0.039 &\nodata &\nodata &\nodata & 4 & 4 & 0 & 0 & 0 &       & 9.9 & 0.5   \\
\enddata
\tablenotetext{a}{X-ray detection significance from ``pwdetect'' (see \citet{sbc04} or \citet{ef06})}

\end{deluxetable}

\begin{deluxetable}{rccccccccl}
\tablecolumns{10}
\tabletypesize{\scriptsize}
%\rotate
\tablecaption{Variables detected in IRAC [3.6] and [4.5] \label{tab_var}}
\tablewidth{0pt}
\tablehead{
\colhead{SST} &
\colhead{[3.6]$_{\rm 2004. Mar. 6}$} & \colhead{[3.6]$_{\rm 2004. Oct. 8}$} &
\colhead{$\Delta$[3.6]} &
\colhead{[4.5]$_{\rm 2004. Mar. 6}$} & \colhead{[4.5]$_{\rm 2004. Oct. 8}$} &
\colhead{$\Delta$[4.5]} &
\colhead{Class} & \colhead{other} & \colhead{Sung et al. (2008)} }
\startdata
 6655 & 11.054 $\pm$ 0.003 & 11.721 $\pm$ 0.034 & -0.667 & 10.520 $\pm$ 0.008 & 11.093 $\pm$ 0.005 & -0.574 & II     & H$\alpha$ & C23131          \\
 7009 & 14.055 $\pm$ 0.025 & 13.477 $\pm$ 0.157 & +0.578 & 13.705 $\pm$ 0.024 & 13.436 $\pm$ 0.012 & +0.269 & II/III &   & C23941          \\
 7252 & 12.823 $\pm$ 0.006 & 13.468 $\pm$ 0.026 & -0.644 & 11.853 $\pm$ 0.037 & 12.461 $\pm$ 0.013 & -0.608 & I      &   &                 \\
 7899 & 12.149 $\pm$ 0.006 & 11.795 $\pm$ 0.001 & +0.354 & 12.082 $\pm$ 0.073 & 11.660 $\pm$ 0.007 & +0.422 & II/III & X+H$\alpha$ & C26021          \\
 8570 & 13.546 $\pm$ 0.093 & 13.085 $\pm$ 0.025 & +0.461 & 13.523 $\pm$ 0.018 & 13.246 $\pm$ 0.027 & +0.277 &        &   & C27455          \\
 9117 &  9.924 $\pm$ 0.054 &  9.548 $\pm$ 0.010 & +0.376 &  9.544 $\pm$ 0.003 &  9.179 $\pm$ 0.010 & +0.365 & II     & X+H$\alpha$ & S1968           \\
 9322 & 11.492 $\pm$ 0.005 & 11.089 $\pm$ 0.009 & +0.403 & 11.122 $\pm$ 0.000 & 10.867 $\pm$ 0.036 & +0.255 & II/III & X+H$\alpha$ & C29082          \\
10455 & 12.518 $\pm$ 0.028 & 12.961 $\pm$ 0.004 & -0.443 & 10.507 $\pm$ 0.012 & 10.890 $\pm$ 0.033 & -0.382 & I      &   &                 \\
11699 & 13.886 $\pm$ 0.022 & 13.076 $\pm$ 0.049 & +0.810 & 12.818 $\pm$ 0.019 & 12.103 $\pm$ 0.007 & +0.715 &        & X &                 \\
11802 & 12.295 $\pm$ 0.002 & 11.754 $\pm$ 0.001 & +0.541 & 10.794 $\pm$ 0.009 & 10.298 $\pm$ 0.020 & +0.496 & I      &   &                 \\
11903 & 12.947 $\pm$ 0.009 & 12.632 $\pm$ 0.003 & +0.315 & 12.360 $\pm$ 0.020 & 11.939 $\pm$ 0.020 & +0.421 & II     &   &                 \\
12051 & 12.329 $\pm$ 0.033 & 11.904 $\pm$ 0.019 & +0.425 & 11.874 $\pm$ 0.031 & 11.470 $\pm$ 0.032 & +0.404 & II     & X+H$\alpha$ & C34222          \\
12722 & 12.252 $\pm$ 0.098 &      \nodata       &\nodata & 10.997 $\pm$ 0.031 & 11.682 $\pm$ 0.002 & -0.685 &        & X &                 \\
13123 & 12.441 $\pm$ 0.003 & 12.990 $\pm$ 0.106 & -0.549 & 11.359 $\pm$ 0.013 & 12.991 $\pm$ 0.088 & -1.632 &        &   &                 \\
13241 & 12.409 $\pm$ 0.004 & 11.719 $\pm$ 0.028 & +0.690 &      \nodata       & 11.287 $\pm$ 0.044 &\nodata &        &   &                 \\
13259 & 13.247 $\pm$ 0.020 & 12.558 $\pm$ 0.001 & +0.688 & 11.733 $\pm$ 0.018 & 11.354 $\pm$ 0.003 & +0.379 &        &   &                 \\
13295 & 11.443 $\pm$ 0.009 & 12.524 $\pm$ 0.056 & -1.080 &      \nodata       &      \nodata       &\nodata &        &   & C36017          \\
13306 & 10.561 $\pm$ 0.003 & 11.287 $\pm$ 0.026 & -0.726 &  9.291 $\pm$ 0.019 &  9.981 $\pm$ 0.039 & -0.690 & I      &   &                 \\
13401 & 14.277 $\pm$ 0.001 & 13.403 $\pm$ 0.049 & +0.874 & 13.366 $\pm$ 0.012 & 12.494 $\pm$ 0.019 & +0.872 &        &   &                 \\
13432 & 12.857 $\pm$ 0.048 & 12.972 $\pm$ 0.118 & -0.115 & 12.059 $\pm$ 0.054 & 13.051 $\pm$ 0.016 & -0.992 &        & X+H$\alpha$ & C36216          \\
13480 & 10.659 $\pm$ 0.027 & 11.105 $\pm$ 0.060 & -0.445 &  9.813 $\pm$ 0.003 & 10.508 $\pm$ 0.002 & -0.695 & II     & X &                 \\
13612 & 12.581 $\pm$ 0.001 & 12.475 $\pm$ 0.027 & +0.106 & 12.295 $\pm$ 0.010 & 11.867 $\pm$ 0.009 & +0.429 & II     &   &                 \\
13715 & 12.217 $\pm$ 0.014 & 12.704 $\pm$ 0.062 & -0.488 &      \nodata       &      \nodata       &\nodata &        & H$\alpha$ & C36610          \\
13734 & 11.143 $\pm$ 0.028 & 11.555 $\pm$ 0.027 & -0.412 & 10.571 $\pm$ 0.030 & 10.886 $\pm$ 0.015 & -0.315 & II     & X+H$\alpha$ & C36644          \\
13749\tablenotemark{a} & 13.144 $\pm$ 0.058 & 12.543 $\pm$ 0.043 & +0.601 & 11.603 $\pm$ 0.095 & 11.090 $\pm$ 0.031 & +0.513 &        & X &                 \\
14514 & 14.474 $\pm$ 0.036 & 13.851 $\pm$ 0.043 & +0.623 & 13.411 $\pm$ 0.021 & 12.810 $\pm$ 0.127 & +0.602 &        &   &                 \\
15726 & 12.173 $\pm$ 0.006 & 11.750 $\pm$ 0.019 & +0.423 & 11.964 $\pm$ 0.041 & 11.283 $\pm$ 0.000 & +0.681 &        & X+H$\alpha$ & C39809+C39817   \\
16833 & 12.796 $\pm$ 0.018 & 12.512 $\pm$ 0.003 & +0.284 & 12.419 $\pm$ 0.001 & 11.951 $\pm$ 0.009 & +0.468 & II     & H$\alpha$ & C41880          \\
\enddata
\tablenotetext{a}{Used only data from long exposure images because short and
long exposure images show large systematic differences probably due to the nearby bright star SST 13808.}
\end{deluxetable}

\begin{deluxetable}{lccccccccccc}
\tablecolumns{10}
\tabletypesize{\scriptsize}
\rotate
\tablecaption{Primordial Disk Fraction of Selected Young Clusters and Stellar Groups \label{tab_disk}}
\tablewidth{0pt}
\tablehead{
\colhead{Cluster} & \colhead{age (Myr)} & \colhead{Sp Type\tablenotemark{a}} & \colhead{disk fraction (\%)} &
\colhead{N$_{\rm total}$\tablenotemark{b}} & \colhead{N$_{\rm LM}$\tablenotemark{b}} & \colhead{N$_{\rm pdisk}$\tablenotemark{b}} & \colhead{mass range} &
\colhead{selection criteria\tablenotemark{c}} & \colhead{PMS model\tablenotemark{d}} & \colhead{reference\tablenotemark{e}} }
\startdata
NGC 1333 & $\leq$ 1 & - & 80.7 $\pm$ 16.0 & $\sim$ $\sim$90 & 57 & 46 & $K_s <$ 12.5 & IR & BCAH98 & 1 \\
Serpens & $\leq$ 1 & - & 74.8 $\pm$ 16.1 & 137 & 49 & 36 & $K_s <$ 12.6 & X, IR &  - & 2 \\
NGC 2244 & 1.7 $\pm$ 0.2 & O5V & 36.2 $\pm$ 4.9 & 748 & 204 & 74 & $\log m =$ 0.0 -- -0.3 & X, IR & SDF00 & 3, 16, 17 \\
$\sigma$ Ori & 1.9 $\pm$ 0.3 & O9.5V & 48.5 $\pm$ 8.5 & 194 & 99 & 48 & $\log m =$ 0.0 -- -0.7 & X, IR, V & SDF00 & 4 \\
NGC 2068/2071 & 2.0 $\pm$ 1.0 & B1.5V & 54.2 $\pm$ 13.2 & 67 & 48 & 26 &  Sp = K7 -- M4 & Sp & SDF00 & 5 \\
IC 348 & 2.5 $\pm$ 0.5 & B5V & 34.3 $\pm$ 6.5 & 307 & 108 & 37 &  Sp = K7 -- M4 & X, IR, Sp & BCAH98 & 6 \\
Tr 37 & 2.6 $\pm$ 0.3 & O6V & 48.5 $\pm$ 7.4 & 172 & 130 & 63 & $\log m =$ 0.0 -- -0.4 & Sp & SDF00 & 7, 15 \\
NGC 2264 S Mon & 3.1 $\pm$ 0.3 & O7V & 26.6 $\pm$ 4.9 & 327 & 139 & 37 & $\log m =$ 0.0 -- -0.6 & X, H$\alpha$, IR, Sp & SDF00 & This\\
$\gamma$ Vel & 3.4 $\pm$ 0.5 & WC & 2.4 $\pm$ 2.5 & 141 & 41 & 1 & $\log m =$ 0.0 -- -0.7 & X, IR & SDF00 & 8 \\
NGC 2362 & 4.0 $\pm$ 1.0 & O9Ib & 7.7 $\pm$ 2.2 & 232 & 168 & 13 & $\log m =$ 0.0 -- -0.7 & X, IR, Sp & BCAH98 & 11 \\
Upp Sco & 5.0            & B8Iab/B0.5V & 14.1 $\pm$ 4.4 & 204 & 85 & 12 & Sp = K7 -- M4 & Sp & - & 12 \\
$\eta$ Cha & 6.0 $\pm$ 2.0 & B8V & 28.6 $\pm$ 22.9 & 18 & 7 & 2 & Sp = K7 -- M4 & Sp & BCAH98 & 9, 10\\
25 Ori & 8.0 $\pm$ 1.5 & B1Vpe & 1.9 $\pm$ 1.9 & 115 & 54 & 1 & Sp = K7 -- M4  & Sp & SDF00 & 13, 14\\
       &               &       &               &     &    &   & or $J$ = 12 -- 13.2 &    &       & \\
NGC 7160 & 10 $\pm$ 2.0 & B1II-III & 7.1 $\pm$ 7.4 & 25 & 14 & 1 & Sp = K7 -- M4 & Sp & SDF00 & 7, 15 \\
\enddata
\tablenotetext{a}{Spectral type of the earliest or evolved star}
\tablenotetext{b}{N$_{\rm total}$ : total number of member stars, N$_{\rm LM}$ : number of low-mass star in the mass range (column 8), N$_{\rm pdisk}$ : number of stars with promordial disks}
\tablenotetext{c}{X: X-ray emission stars, IR: IR excess stars from Spitzer observation, H$\alpha$: H$\alpha$ photometry, Sp: Spectroscopically confirmed members}
\tablenotetext{d}{BCAH98: \citet{bcah98}, SDF00: \citet{sdf00} }
\tablenotetext{e}{1: \citet{rag08}, 2: \citet{we07}, 3: \citet{wj08}, 4: \citet{hj07a}, 5: \citet{fm08}, 6: \citet{cjl06}, 7: \citet{saa06}, 8: \citet{hj08}, 9: \citet{mhlf05},
                  10: \citet{ls04}, 11: \citet{dh07}, 12: \citet{cmhm06}, 13: \citet{hj07b}, 14: \citet{bc07}, 15: \citet{saa05}, 16: \citet{zb07}, 17: \citet{pns02} }
\end{deluxetable}

\begin{deluxetable}{rccccccccccc}
\tablecolumns{10}
\tabletypesize{\scriptsize}
%\rotate
\tablecaption{New BMS star candidates \label{tab_bms}}
\tablewidth{0pt}
\tablehead{
\colhead{SST} & \colhead{Sung et a. (2008)} &
\colhead{[3.6]} & \colhead{[4.5]} & \colhead{[5.8]} & \colhead{[8.0]} & \colhead{[24]} &
\colhead{H$\alpha$ emission} & \colhead{IR Class} & \colhead{SED slope} & \colhead{Q$_{CC}$} & \colhead{$\Delta I$\tablenotemark{a}} }
\startdata
 2233 & C12598 & 11.377 & 10.701 & 10.245 &  9.480 & 6.483 & yes & II & -0.58 $\pm$ 0.03 & 2.0 & 0.24 \\
 6655 & C23131 & 11.377 & 10.814 & 10.355 &  9.703 & 6.778 & yes & II & -0.69 $\pm$ 0.09 & 2.0 & 0.45 \\
 6892 & C23663 & 13.867 & 13.286 & 12.683 & 11.759 & 8.431 & yes & II & -0.33 $\pm$ 0.13 & 2.0 & 4.19 \\
 7361 & C24781 & 15.739 & 14.712 & 13.378 & 11.830 & 7.603 & no  & I  & +1.35 $\pm$ 0.47 & 1.0 & 2.50 \\
 8261 & C26827 & 13.277 & 12.549 & 12.063 & 11.386 & 8.572 & yes & II & -0.70 $\pm$ 0.08 & 2.0 & 2.75 \\
10184\tablenotemark{b} & C30920+C30962 & 13.434 & 13.302 & 13.030 & 10.598 & 7.672 & no & II & -0.04 $\pm$ 0.50 & 2.0 & 0.51 \\
10710\tablenotemark{c} & C32005 & 11.773 & 11.091 &  9.556 &  7.853 & 3.135 & ? & I & 1.34 $\pm$ 0.10 & 1.0 & 2.58 \\
13708 & C36601 & 14.512 & 13.759 & 13.048 & 12.225 &       & yes & II & -0.19 $\pm$ 0.11 & 2.0 & 11.62 \\
18887 & C46626 & 14.028 & 13.121 & 12.145 & 11.107 & 7.900 & no  & I  & +0.09 $\pm$ 0.28 & 1.3 & 3.93 \\
21725 & C54567 & 12.666 & 12.086 & 11.434 & 10.716 & 7.702 & yes & II & -0.55 $\pm$ 0.03 & 2.0 & 0.30 \\
\enddata
\tablenotetext{a}{Magnitude difference from the faint limit of the PMS locus in NGC 2264}
\tablenotetext{b}{If the actual optical counterpart of SST 10184 is C30962, SST 10184 may be a star-forming galaxy with PAH emission. See \S 2.5.}
\tablenotetext{c}{\citet{ko84} noted as an H$\alpha$ emission star, but no signature of H$\alpha$ emission from CCD photometry was detected.}
\end{deluxetable}

\end{document}